\documentstyle[preprint,aps,prb,epsf]{revtex}
\begin{document}
\preprint
\widetext
\title{Strong-coupling expansions for the anharmonic Holstein model
and for the Holstein-Hubbard model}
\author{J.\ K.\ Freericks$^1$ and G.~D. Mahan$^2$}
\address{$^1$Department of Physics, Georgetown University, Washington, DC 
20057--0995\\
$^2$Department of Physics and Astronomy, University of Tennessee, Knoxville,
Tennessee 37996--1200
and Solid State Division, Oak Ridge National Laboratory, P. O. Box 2008,
Oak Ridge, Tennessee 37831--6030\\}
\date{\today}
\maketitle
\widetext
\begin{abstract}
A strong-coupling expansion is applied to the anharmonic Holstein model
and to the Holstein-Hubbard model through fourth order in the hopping
matrix element.  Mean-field theory is then employed to determine transition
temperatures of the effective (pseudospin) Hamiltonian.  We find that 
anharmonic effects are not easily mimicked by an on-site Coulomb repulsion,
and that anharmonicity strongly favors superconductivity relative to 
charge-density-wave order.  Surprisingly, the phase diagram is strongly
modified by relatively small values of the anharmonicity.
\end{abstract}
\renewcommand{\thefootnote}{\copyright}
\footnotetext{ 1996 by the authors.  Reproduction of this article by any means
is permitted for non-commercial purposes.}
\renewcommand{\thefootnote}{\alpha{footnote}}


\section{Introduction}

The interaction of the conduction electrons in a solid with the lattice
vibrations is described by the so-called
electron-phonon problem.  Migdal\cite{migdal} and
Eliashberg\cite{eliashberg} pioneered the study of such interacting
fermion-boson systems in the limit where the phonons are all harmonic,
and the phonon energy scale is much smaller than the electronic energy scale.
Vertex corrections can be neglected in this case\cite{migdal} and a 
self-consistent theory can be constructed that is exact in the limit of 
weak-coupling; the theory is an expansion in powers of the coupling strength
multiplied by the ratio of the phonon energy scale to the electronic 
energy scale.

But phonons in real materials are never purely harmonic---higher-order 
(anharmonic) contributions to the phonon potential are always present.  
This phonon anharmonicity is
responsible for many different physical effects in solids.  For example,
thermal expansion arises purely from anharmonic effects---a harmonic crystal
does not change its volume upon heating. Such anharmonic effects have been
treated in an approximate fashion:  in the quasiharmonic 
approximation\cite{liebfried_ludwig} only the effect of thermal expansion
is taken into account by postulating that the phonon frequencies have a 
dependence upon the volume of the crystal, as described by the Gr\" uneisen
parameter; in the self-consistent harmonic approximation\cite{horner}
the harmonic force constants are self-consistently replaced by their
thermal averages over all possible motions of the other atoms---it is
used to describe systems with strong anharmonicity; finally, in the
pseudoharmonic approximation\cite{reissland} both thermal expansion
effects and phonon-phonon interactions are taken into account by employing
the quasiharmonic approximation plus a perturbative expansion in the 
phonon-phonon interaction.  Superconducting transition temperatures have
also been studied, with the result that anharmonicity does not enhance
the transition temperature in the weak to moderate coupling regime where
Eliashberg theory applies\cite{sofo_mahan}.
But an exact treatment of lattice anharmonicity is difficult from a 
theoretical point of view because an anharmonic ``perturbation''
is never a small perturbation; the phonon wavefunctions are always dominated by
the anharmonic terms in the potential as the phonon coordinate becomes large.
Even the qualitative effects of lattice anharmonicity on superconductivity
are not well understood.  All that is known rigorously about the anharmonic
electron-phonon problem is that the ground state must contain a spin
singlet\cite{freericks_lieb} for even numbers of electrons on a finite lattice.

Much progress can be made however, in the limit of strong coupling (the
electron-phonon interaction is much larger than the hopping integral of
the electrons), where the electrons strongly bind together into preformed 
pairs (called bipolarons) and the ground state of the system is highly 
degenerate. Degenerate perturbation theory (in the kinetic energy of the
electrons) about this bipolaronic ground state produces an expansion
in inverse powers of the coupling strength.  This theory has been 
exhaustively analyzed to second order in the hopping\cite{bipolarons}
and has recently been studied to fourth order in the 
hopping\cite{freericks_strong}.  In this contribution, the fourth-order
calculations are extended to include both anharmonic phonons and a direct
electron-electron repulsion.

The simplest electron-phonon model that includes both anharmonic effects
and direct electron-electron repulsion is the anharmonic Holstein-Hubbard
model\cite{holstein,hubbard} in which the conduction electrons interact
with themselves and with local phonon modes:
\begin{eqnarray}
H&=&-  \sum_{i,j,\sigma} t_{ij} c_{i\sigma}^{\dag }c_{j\sigma}
+ \sum_i ( g x_i - \mu )(n_{i\uparrow}+n_{i\downarrow}) + U_c\sum_i n_{i
\uparrow}n_{i\downarrow}\cr
&+&\frac{1}{2M}\sum_i p_{i}^2+\frac{1}{2}M\Omega^2\sum_ix_i^2+\alpha_{anh}\sum_i
x_i^4\quad ,
\label{eq: ham}
\end{eqnarray}
where $c_{i\sigma}^{\dag}$ ($c_{i\sigma}$) creates (destroys) an electron at
site $i$ with spin $\sigma$, $n_{i\sigma}=c_{i\sigma}^{\dag}c_{i\sigma}$ is
the electron number operator, and $x_i$ ($p_i$) is the phonon coordinate
(momentum) at site $i$.  The hopping of electrons
between lattice sites $i$ and $j$ is governed by the hopping matrix element
$t_{ij}$ ($t_{ij}$ is a Hermitian matrix).

The local phonon has a mass $M$ and a
frequency $\Omega$ associated with it; the combination 
$\kappa := M\Omega^2$  is a
spring constant that measures the stored energy per unit length squared
in the phonon coordinate.  The anharmonic contribution to the phonon potential
energy is chosen to be a quartic in the phonon coordinate with a strength 
$\alpha_{anh}$.  The electron-phonon interaction strength is
parameterized by an energy per unit length and is denoted $g$.  A useful
combination of fundamental parameters is the bipolaron binding energy
(in the harmonic limit with $U_c=0$)
\begin{equation}
U := - \frac{g^2}{M\Omega^2}=-\frac{g^2}{\kappa} \quad ,
\label{eq: udef}
\end{equation}
which determines the energy scale for the effective electron-electron
interaction mediated by the phonons.  This attraction competes with the direct
Coulomb repulsion denoted by $U_c$.  The chemical potential is $\mu$.

The hopping matrix elements $t_{ij}$ are used to define the energy scale.  The
choice $t_{1NN}=:t^*/2\sqrt{d}$ with $t^*=1$ and $d$ the dimensionality
of the lattice, is made so as to have a well-defined limit when $d\rightarrow
\infty$.  The
mass is then set equal to one $(M=1)$ leaving $U$, $U_c$, $\alpha_{anh}$, and 
$\Omega$ as free parameters.  The strong-coupling expansion is a
perturbative expansion in the hopping terms of Eq.~(\ref{eq: ham}) and is
valid when the bipolaron binding energy is much larger than the electronic
hopping integral.  Another important parameter in the strong-coupling limit
is the polaron band-narrowing parameter denoted by $S=|U|/\Omega$.

The original Holstein Hamiltonian corresponds to the case $\alpha_{anh}=U_c=0$.
Both the harmonic Holstein model and the harmonic Holstein-Hubbard model
have been solved exactly in the limit of infinite dimensions via quantum
Monte Carlo simulation\cite{freericks_qmc_holst,freericks_qmc_holsthubb}.
These models display charge-density-wave (CDW) order near half filling and 
superconductivity (SC) away from half-filling.  As the phonon frequency is 
increased, the SC is favored relative to the CDW order.  However, in the
strong-coupling limit, CDW order is favored over SC because of the 
band-narrowing effect of the bipolaron.  The quantum Monte Carlo simulations
also found that the effective phonon potential (determined after integrating
out the effects of the electrons) generically acquires a double-well structure
signifying the formation of a bipolaron, and indicating that a strong-coupling
expansion should be accurate even down to moderate values of the coupling
strength.  This has proven to be true for the harmonic 
case\cite{freericks_strong} and is likely to also hold over a more restricted
region for the anharmonic case.

An initial analysis of the anharmonic model in the strong-coupling 
limit ($t_{ij}=0$) can be made by using the Born-Oppenheimer
approximation\cite{hirsch_anharm}:  the phonon frequency is assumed to be 
smaller than any
of the other energy scales so the phonons can be approximated by static lattice
distortions corresponding to the minimum of the phonon potential energy.
Since the phonons couple linearly to the electronic charge,
the equilibrium phonon coordinate varies when there are zero, one, or two 
electrons on a site.  The origin $x_0=0$ is chosen to correspond to the case
with no electrons  on a site.  Then $x_1$ and $x_2$
denote the equilibrium coordinates with one or two electrons on a lattice site.
In the harmonic case the relative distances $x_1-x_0$ and $x_2-x_1$ are 
identical, which is a requirement for particle-hole symmetry.  When a
lattice anharmonicity is turned on, the equilibrium phonon coordinates with
one and two electrons on a lattice site all move toward the origin, but the
relative distances are no longer symmetric, rather the distance $x_2-x_1$ 
becomes significantly smaller than $x_1-x_0$, as can be seen by calculating
the perturbative shift in the equilibrium coordinates\cite{hirsch_anharm} 
as a function of the anharmonicity $\alpha_{anh}$
\begin{equation}
x_0=0\quad ,\quad x_1=-\frac{g}{M\Omega^2}+\frac{4\alpha_{anh} g^3}{M^4\Omega^8}
\quad ,\quad x_2=-\frac{2g}{M\Omega^2}+\frac{32\alpha_{anh} g^3}{M^4\Omega^8}
\quad .
\label{eq: bo_equilib}
\end{equation}

This asymmetric shift of the equilibrium phonon coordinate as a function of
lattice anharmonicity causes two main effects: (1) the model loses 
particle-hole symmetry which allows a new type of superconductivity 
to emerge \cite{hirsch_anharm} and
(2) the effective electron-electron attraction is sharply reduced as can
be seen by a plot of the bipolaron binding energy in Figure 2~(b).  Thus the
lattice anharmonicity generates an effective {\it retarded} repulsive 
interaction between the electrons and breaks particle-hole symmetry, removing 
the nesting instability of the CDW at half filling and weak
coupling.  One expects that the lattice anharmonicity thereby to favor 
SC relative to CDW order, although it is also likely that anharmonic effects
will reduce the transition temperatures (except close
to the filled band, where the new hole-superconductivity mechanism can take
over).  What is surprising is that a rather small lattice anharmonicity can
have a large effect on the electron-phonon problem.  

Since the main effects of lattice anharmonicities are driven by the asymmetric
distribution
of the equilibrium phonon coordinate when there is zero, one, or two electrons
on a lattice site, one expects that anharmonic effects will be strongest in
the small-phonon-frequency limit.  This is because the phonon coordinates
all approach zero in the high-frequency limit (because the phonon reacts
instantaneously to the change in the electrons)
and these asymmetric effects disappear.  Since phonon frequencies tend to be
small in real materials, anharmonic effects can be important even if the
phonon potential energy appears to be well approximated by a harmonic potential,
i.~e., if $\alpha_{anh}$ is small.

Hirsch's new mechanism for superconductivity arises from an examination of the
anharmonic electron-phonon model in the static limit\cite{hirsch_anharm}.
The tunneling matrix element for a polaron from one
lattice site to its nearest neighbor depends exponentially on the difference
$x_1-x_0$ if there is no electron on the neighboring site and exponentially on
$x_2-x_1$ if there is an electron occupying the neighboring site.  The 
exponential dependence arises from the Franck-Condon overlap factors.
Since these
two values can be significantly different in an anharmonic model, one finds
that the electronic motion will be dominated by a kinetic energy that
depends on the density of the electrons at a given site.  This is precisely
the physical situation needed to generate superconductivity from kinetic
energy effects---paired electrons have a lower effective mass than a
single electron, which induces the superconductivity transition at a low enough
temperature.  This novel mechanism for generating superconductivity has been
analyzed via weak-coupling mean-field theory analyses\cite{hirsch_hole}
but has not yet been shown to exist in the exact solution of any model system.
This new mechanism for superconductivity disappears, however, in the 
strong-coupling limit where restriction is made to consider only empty sites
and bipolarons.  The single-electron (polaronic) states are integrated out
because the bipolaron binding energy is much larger than the electronic
hopping integral.  Electron-hole symmetry is restored since these polaronic
states only appear in virtual processes.

The anharmonicity has a much different effect than a direct electron-electron
repulsion.  The electron-electron repulsion uniformly reduces the
bipolaron binding energy without changing the equilibrium phonon coordinates
when there are zero, one, or two electrons at a lattice site.  Thus, (1) the
bipolaron binding energy can become negative (signifying there is no 
electron-electron pairing) and (2)
the system explicitly retains it's electron-hole
symmetry.  Furthermore, since the repulsion is instantaneous, the retardation
effects are unchanged from the case without Coulomb repulsion, i.~e.,
the Franck-Condon overlaps remain the same.  Thus we
expect CDW ordering to always survive at half filling (if there is a net 
electron-electron attraction), and that both SC and
CDW order will disappear as the Coulomb repulsion becomes too large, but
it is not clear whether or not they disappear together, or at different
values of the electron-phonon coupling.

This contribution is a continuation of the work of one of the 
authors\cite{freericks_strong} on the strong-coupling expansion for the
harmonic Holstein model to include both anharmonic and Coulomb repulsion
effects.  In Sec. II, the formalism for the perturbation theory, and
the generation of the effective pseudospin Hamiltonian will be described.
In Sec. III, a mean-field-theory analysis of the pseudospin Hamiltonian will
be given, and appropriate phase diagrams calculated for both the anharmonic
case and the Coulomb repulsion case.  Conclusions and a discussion will follow
in Sec. IV.

\section{Formalism for the Perturbative Analysis}

The strong-coupling expansions are carried out with a method based on 
perturbation theory.  The ground state is a bipolaronic state consisting of
either paired electrons or empty sites.  The distribution of these bipolarons
is not determined to zeroth order in the electronic kinetic energy, so the
ground state is highly degenerate.  The effective Hamiltonian (within this
degenerate subspace) can be determined by using operator methods.
In the late 1950's, Anderson\cite{anderson}  first used such methods to
show that the strong-interaction-strength limit of the Hubbard model
is described by a Heisenberg antiferromagnet with an exchange integral
$j=4|t_{ij}|^2/|U|$ that vanishes as the interaction strength increases.
Kato\cite{kato} described how to determine the effective Hamiltonian for an 
arbitrary degenerate subspace using perturbation theory and
operator methods.  His analysis was applied to
the Hubbard model in one dimension by Klein and Seitz\cite{kleinandseitz}
and in arbitrary dimensions by Takahashi\cite{takahashi} (see also the recent
work by van Dongen\cite{vandongen}).  Beni, Pincus,
and Kanamori\cite{beni} and Hirsch and 
Fradkin\cite{fradkin_hirsch} applied the same methods to the harmonic Holstein
model determining the effective Hamiltonian to second order in the hopping.
This analysis was extended to fourth order by Freericks\cite{freericks_strong}.

Kato's method\cite{kato} begins with a Hamiltonian $H=H_0+T$ with $H_0$ the 
unperturbed Hamiltonian and $T$ the perturbation.  In our case,
$H_0$ corresponds
to the Hamiltonian in Eq.~(\ref{eq: ham}) with $t_{ij}=0$, and the perturbation
$T$ is the electronic kinetic energy.  The ground-state energy is $E_0$,
$Q_0$ denotes the subspace that contains
all of the degenerate ground states, and the projection
operator onto $Q_0$ is $P_0$:
\begin{equation}
H_0P_0=P_0H_0=E_0P_0\quad\quad,\quad\quad P_0^2=P_0\quad .
\label{eq: p0def}
\end{equation}
As the perturbation is turned on, the eigenstates
will evolve into a new subspace $Q$ with corresponding
projection operator $P$.  If it is
assumed that the subspace $Q$ has a nonzero overlap with the unperturbed
subspace $Q_0$, then the
standard eigenvalue equation $(H-E)|E\rangle =0$ can be projected
onto the unperturbed subspace $Q_0$: $P_0(H-E)PP_0|E\rangle =0$, 
to yield an effective equation for the perturbed eigenvalue $E$.
The Hamiltonian $P_0HPP_0$ acts purely within the unperturbed subspace 
$Q_0$ and has an overlap operator $P_0PP_0$ that is not equal to the identity.
Taking into account this nontrivial overlap, results in an effective Hamiltonian
of the form\cite{lowdin,freericks_strong} $H_{eff}=H_0+H_2+H_4+\ldots$, since
only even powers of the perturbation enter for the generalized Holstein model.
The first two nontrivial terms of the effective Hamiltonian satisfy
\begin{equation}
H_2:= P_0T\frac{1-P_0}{E_0-H_0}TP_0\quad,
\label{eq: h2def}
\end{equation}
and
\begin{eqnarray}
H_4&:=& P_0T\frac{1-P_0}{E_0-H_0}T\frac{1-P_0}{E_0-H_0}
T\frac{1-P_0}{E_0-H_0}TP_0\cr
&-&\frac{1}{2}\left [ P_0T\frac{1-P_0}{(E_0-H_0)^2}TP_0T\frac{1-P_0}{E_0-H_0}
TP_0+ P_0T\frac{1-P_0}{E_0-H_0}T\frac{1-P_0}{(E_0-H_0)^2}TP_0\right ]\quad.
\label{eq: h4def}
\end{eqnarray}

The expansion for the effective Hamiltonian can be expressed graphically by
a set of diagrams.  A solid line denotes virtual processes where an electron
hops from site $i$ to site $j$ with strength $t_{ij}$.  All diagrams must be
closed, since the effective Hamiltonian acts solely
within the degenerate subspace $Q_0$ implying each virtually
broken electron pair must be
restored.  There is only one possibility for the second-order term, which 
corresponds to either hopping from site $i$ to site $j$
and hopping back to site $i$, or which corresponds to subsequent hops from 
site $i$ to site $j$.  The diagram that illustrates both of these processes is 
depicted in Figure~1(a).  There are four possible diagrams that 
contribute to fourth order which are also depicted in Figure 1.  The first three
diagrams are linked diagrams which form nonvanishing contributions to
the effective Hamiltonian.  The last diagram 1(e) is an unlinked diagram which
does not contribute to the effective Hamiltonian because the contributions
from the positive and negative terms in Eq.~(\ref{eq: h4def}) cancel.  The 
unlinked diagrams must cancel in order to have an energy per lattice site that 
is finite in the thermodynamic limit.  The nonvanishing
fourth-order terms fall into three
categories: those that link two distinct sites [Figure~1(b)]; three distinct
sites [Figure~1(c)]; or four distinct sites [Figure~1(d)].

The matrix elements for the effective Hamiltonian are determined by introducing
appropriate complete sets of states between each of the operator
factors in Eqs.~(\ref{eq: h2def}) and (\ref{eq: h4def}) for each of the 
possible intermediate virtual states summarized by the diagrams in Figure~1.
In the following section, thermodynamic phase transitions are determined
from the effective Hamiltonian via a mean-field-theory analysis.  An 
approximation is made here, that the transition temperature $T_c$ is much 
smaller than the excitation energy of the lowest state above the ground
state for the local anharmonic phonon with zero or two electrons
on that site.  In the harmonic case $(\alpha_{anh}=0)$, this excitation energy
is $\Omega$, and even in the anharmonic case, this excitation energy
remains of the order of magnitude of $\Omega$.  If the transition temperature
is much smaller than the excitation energy, then restriction can be made to
the degenerate subspace $Q_0$ corresponding to the ground state of the phonons
and one need not consider the effective Hamiltonian in the subspaces
corresponding to excited phonon states.  This is not a restrictive 
approximation, because the small-frequency limit is already known to be
singular, since the degenerate subspace $Q_0$ becomes much larger 
when $\Omega=0$ than when $\Omega\ne 0$.  $\Omega/t^*$ will be set equal to 0.5
for the numerical work in Section III.

Determination of the effective Hamiltonian proceeds in a similar fashion to
the harmonic case\cite{beni,fradkin_hirsch,freericks_strong}.  The minimum
of the anharmonic potential lies at $x_0$, $x_1$, and $x_2$ when there are
zero, one, or two electrons, respectively, at a given lattice site (in the
harmonic case we have $x_n=-ng/M\Omega^2$).  Let $|+m\rangle$, $|m\rangle$, and $|-m\rangle$ denote the $m$th
anharmonic oscillator state centered about the origin with zero, one, and
two electrons respectively, and $E_+(m)$, $E(m)$, and $E_-(m)$ denote the
corresponding eigenvalues under the unperturbed Hamiltonian $H_0$.  
The overlaps $\langle\pm m|n\rangle$ will need to be calculated numerically
in the general case.  For the harmonic oscillator, a simple form is found
\begin{equation}
\langle\pm m|n\rangle =\frac{1}{\sqrt{m!n!}} e^{-\frac{1}{4}S}\langle 0|[a\pm
\sqrt{S/2}]^m [a^{\dag}\mp\sqrt{S/2}]^n|0\rangle \quad.
\label{eq: overlap_harm}
\end{equation}
Here $a^{\dag}$ ($a$) is the harmonic-oscillator creation (annihilation) 
operator.  In the harmonic case, one can use Eq.~(\ref{eq: overlap_harm})
to determine the parameters of the
effective Hamiltonian analytically. 

We will concentrate, however, on the anharmonic case, where the local phonon
problem must be solved numerically.  The unperturbed Hamiltonian $H_0$ consists
of a collection of local phonon Hamiltonians, one for each lattice site,
with a fixed number of electrons $n=0,1,2$ at each lattice site $i$.  The
local Hamiltonian is
\begin{equation}
H_{loc}=\frac{1}{2M}p^2+\frac{1}{2}M\Omega^2x^2+\alpha_{anh}x^4+gnx\quad ,
\label{eq: hloc}
\end{equation}
where we ignore terms that do not depend on the phonon coordinate or momentum.
For harmonic phonons, the unit of length is (Planck's constant is set equal
to one) $x^*:=1/\sqrt{M\Omega}$.  Reexpressing the Hamiltonian in terms of
a dimensionless distance $y:=x/x^*$ yields
\begin{equation}
H_{loc}=\frac{\Omega}{2}\bar H(n,y)\quad ,\quad \bar H(n,y):=-\frac{d^2}{dy^2}
+\bar V(n,y)\quad ,\quad \bar V(n,y):=y^2+vy^4+wny\quad ,
\label{eq: hloc2}
\end{equation}
with $v:=2\alpha_{anh}/M^2\Omega^2$ and $w:=2g/\sqrt{m\Omega^3}$.  The
Schr\" odinger equation $\bar H\phi_m=e(m)\phi_m$ is then solved numerically
with the Numerov algorithm\cite{mahan_num}.  Schr\" odinger's equation is cast
into a three term recursion relation which is iterated from the far left
and from the far right, and matched at the middle.  To start the iteration
requires a good initial guess for the eigenvalue.  This is provided by the
WKBJ guess $e_{WKBJ}(m)$ found from
\begin{equation}
2m+1=\frac{2}{\pi}\int dy\sqrt{e_{WKBJ}(m)-\bar V(y)}\quad .
\label{eq: wkbj}
\end{equation}
We generally solve for the thirty lowest eigenvalues and eigenvectors in the
sectors with 0, 1, or 2 electrons.  

The effective Hamiltonian is determined by evaluating all possible contributions
from each of the nonvanishing diagrams in Figure 1.  Consider first the 
second-order term in Figure~1(a).  If both site $i$ and site $j$ are occupied
by bipolarons, then the hopping matrix cannot connect the two sites by the
Pauli principle.  Similarly, if both sites are empty, there is no connection
by the hopping term.  It is only if one site is occupied by a bipolaron and
the other site empty, that the hopping matrix can connect through a virtual
state back to the degenerate subspace.  Consider the case with a bipolaron
$\uparrow\downarrow$ at site $i$ and an empty site $0$ at site $j$.  The
hopping perturbation breaks the electron pair, with one electron hopping
to site $j$ and either hopping back to site $i$ or the other electron
also hops from site $i$ to site $j$.  Both processes are illustrated
schematically below (with $\uparrow$ or $\downarrow$ corresponding
to a single electron at a lattice site):
\begin{equation}
\uparrow\downarrow\quad 0 \quad \quad \Rightarrow\quad\quad \uparrow\quad
\downarrow\quad\quad\Rightarrow\quad\quad\uparrow\downarrow\quad 0\quad,
\label{eq: 2ndordera}
\end{equation}
\begin{equation}
\uparrow\downarrow\quad 0 \quad \quad \Rightarrow\quad\quad \uparrow\quad
\downarrow\quad\quad\Rightarrow\quad\quad0\quad\uparrow\downarrow\quad,
\label{eq: 2ndorderb}
\end{equation}
where only one of the two possible intermediate states is shown.

It is convenient to express the effective Hamiltonian in terms of pseudospin
operators\cite{liebyangzhang}.  If the lattice is bipartite, so that it can
be separated into $A$ and $B$ sublattices with nonzero hopping matrix
elements only between sublattices $A$ and $B$, then one can define 
pseudospin operators via
\begin{equation}
J_j^+:=(-1)^j c_{j\uparrow}^{\dag}c_{j\downarrow}^{\dag}\quad,\quad
J_j^-:=(J_j^+)^{\dag}\quad,\quad J_j^z:=\frac{1}{2}[n_{j\uparrow}+
n_{j\downarrow}-1]\quad,
\label{eq: jdef}
\end{equation}
and the factor $(-1)^j$ is 1 for the $A$ sublattice and $(-1)$ for the
$B$ sublattice.
The pseudospin operators satisfy an SU(2) algebra and form a spin-$\frac{1}{2}$
representation in the strong-coupling limit.  A doubly occupied site 
corresponds to an up spin
and an empty site corresponds to a down spin.  The matrix elements of the
effective Hamiltonian (that connects site $i$ to site $j$) satisfy
\begin{eqnarray}
H_2(i,j)|\uparrow\uparrow\rangle&=&0\quad,\cr
H_2(i,j)|\uparrow\downarrow\rangle&=&
-\frac{1}{2}j_{\parallel}^{(2)}(i,j)|\uparrow\downarrow\rangle
+\frac{1}{2}j_{\perp}^{(2)}(i,j)|\downarrow\uparrow
\rangle\quad , \cr
H_2(i,j)|\downarrow\uparrow\rangle&=&
-\frac{1}{2}j_{\parallel}^{(2)}(i,j)|\downarrow\uparrow\rangle
+\frac{1}{2}j_{\perp}^{(2)}(i,j)|\uparrow\downarrow
\rangle\quad , \cr
H_2(i,j)|\downarrow\downarrow\rangle&=&0\quad,
\label{eq: h2_pseudo}
\end{eqnarray}
which is an $XXZ$ Heisenberg antiferromagnet
\begin{equation}
H_2=\frac{1}{2}\sum_{i,j}\{ j_{\perp}^{(2)}(i,j)\frac{1}{2}
[J_i^+J_j^-+J_i^-J_j^+]+j_{\parallel}^{(2)}(i,j)
[J_i^zJ_j^z-\frac{1}{4}]\} \quad .
\label{eq: h2final}
\end{equation}
Note that the summation is not restricted to $i<j$ (the overall factor of 
$\frac{1}{2}$
is introduced to compensate for double counting).  Be careful not to confuse
the pseudospin Hamiltonian from the original Hamiltonian that consists of 
bipolarons and empty sites.

The two parameters $j_{\parallel}^{(2)}$ and $j_{\perp}^{(2)}$ are determined
by introducing complete sets of states into Eq.~(\ref{eq: h2def}) for each
of the virtual processes in Eqs.~(\ref{eq: 2ndordera}) and (\ref{eq: 2ndorderb})
and employing
the definitions given in Eq.~(\ref{eq: h2_pseudo}).  The results are
\begin{equation}
j_{\parallel}^{(2)}=\sum_{m,n=0}^{\infty}4t_{ij}^2\frac{
\langle -0|m\rangle\langle m|-0\rangle\langle +0|n\rangle
\langle n|+0\rangle}{E_+(0)+E_-(0)-E(m)-E(n)}\quad ,
\label{eq: jpardef}
\end{equation}
\begin{equation}
j_{\perp}^{(2)}=\sum_{m,n=0}^{\infty}4t_{ij}^2\frac{
\langle -0|m\rangle\langle m|+0\rangle\langle +0|n\rangle
\langle n|-0\rangle}{E_+(0)+E_-(0)-E(m)-E(n)}\quad .
\label{eq: jperdef}
\end{equation}

The fourth-order terms are all evaluated in a similar fashion.  They can be
separated into three different forms $H_4=H_4(b)+H_4(c)+H_4(d)$ corresponding
to the three different linked diagrams in Figure~1.  The effective Hamiltonian
for each of these three cases takes the form
\begin{equation}
H_4(b)=\frac{1}{2}\sum_{i,j}\{ j_{\perp}^{(4)}(i,j)\frac{1}{2}
[J_i^+J_j^-+J_i^-J_j^+]+j_{\parallel}^{(4)}(i,j)
[J_i^zJ_j^z-\frac{1}{4}]\} \quad .
\label{eq: h4bfinal}
\end{equation}
\begin{eqnarray}
H_4(c)&=&\frac{1}{2}\sum_{i,j,k}^{\quad} \phantom{}^{\prime}\biggl \{ 
j_{\perp}^{\prime}\frac{1}{2}[J_i^+J_j^-
+J_i^-J_j^++J_k^+J_j^-+J_k^-J_j^+]-j_{\parallel}^{\prime}[J_i^zJ_j^z+
J_k^zJ_j^z-\frac{1}{2}]\cr
&+&j_{\perp}^{\prime\prime}\frac{1}{2}[J_i^+J_k^-+J_i^-J_k^+]+
[j_{\parallel}^{\prime}+j_{\parallel}^{\prime\prime}][J_i^zJ_k^z-\frac{1}{4}] 
\biggr \} \quad ,
\label{eq: h4cfinal}
\end{eqnarray}
\begin{eqnarray}
H_4(d)&=&\frac{1}{8}\sum_{i,j,k,l}^{\quad} \phantom{}^{\prime}\biggl \{ 
\frac{\alpha}{2} +\frac{\delta}{4} +\frac{\nu}{8} -\frac{\nu}{2}
\big [J_i^zJ_j^z+J_i^zJ_l^z+J_k^zJ_j^z+J_k^zJ_l^z \big ]\cr
&+&\frac{\beta+\epsilon}{2}\big [ J_i^+J_j^-+J_i^-J_j^++J_i^+J_l^-+J_i^-J_l^+
+J_k^+J_j^-+J_k^-J_j^++J_k^+J_l^-+J_k^-J_l^+ \big ] \cr
&-&(\delta-\frac{\nu}{2})\big [ J_i^zJ_k^z+J_j^zJ_l^z \big ] +
\frac{\gamma +\mu}{2}\big [ J_i^+J_k^-+J_i^-J_k^++J_j^+J_l^-+J_j^-J_l^+ \big ]
\cr
&+&2(\beta -\epsilon )\big [ J_i^zJ_j^z(J_k^+J_l^-+J_k^-J_l^+)+J_k^zJ_l^z
(J_i^+J_j^-+J_i^-J_j^+)\cr
&\quad&\quad\quad+J_i^zJ_l^z(J_k^+J_j^-+J_k^-J_j^+)+J_k^zJ_j^z
(J_i^+J_l^-+J_i^-J_l^+) \big ] \cr
&+&2(\gamma -\mu )\big [ J_i^zJ_k^z(J_j^+J_l^-+J_j^-J_l^+)+
J_j^zJ_l^z(J_i^+J_k^-+J_i^-J_k^+) \big ] \cr
&+& \frac{\rho}{2}\big [ (J_i^+J_j^-+J_i^-J_j^+)(J_k^+J_l^-+J_k^-J_l^+)+
(J_i^+J_l^-+J_i^-J_l^+)(J_k^+J_j^-+J_k^-J_j^+)\cr
&\quad&\quad -(J_i^+J_k^-+J_i^-J_k^+)(J_j^+J_l^-+J_j^-J_l^+) \big ]
+[-8\alpha +4\delta +2\nu ]J_i^zJ_j^zJ_k^zJ_l^z \biggl \} \quad ,
\label{eq: h4dfinal}
\end{eqnarray}
where the lattice-site-index dependence of the parameters has been suppressed,
and the prime on the summations means that the sites $i,j,k$ and $i,j,k,l$
are all distinct (the overall factors of $\frac{1}{2}$ and $\frac{1}{8}$
are introduced to compensate for double counting).  Explicit expressions for 
each of the parameters in 
Eqs.~(\ref{eq: h4bfinal}), (\ref{eq: h4cfinal}), and (\ref{eq: h4dfinal})
are given in the appendix.

The effective pseudospin Hamiltonian is an anisotropic frustrated 
antiferromagnetic Heisenberg model with additional quartic spin-spin 
interactions.

The effect of the Coulomb repulsion is almost trivial.  The Coulomb repulsion
does not change any of the matrix elements, all it does is shift the energy
of the bipolaron upward by $U_c$: $E_-(m)\rightarrow E_-(m)+U_c$, so the
effect of Coulomb repulsion can be included without much extra effort.
One might try to approximate the effect of the anharmonicity  
by an instantaneous Coulomb repulsion, chosen to match the reduction of the
bipolaron binding energy when the anharmonicity is included, i.~e. define
\begin{equation}
U_c(\alpha_{anh}):=E_-(0;\alpha_{anh})+E_+(0;\alpha_{anh})-2E(0;\alpha_{anh})
+|U|
\quad ,
\label{eq: ucalpha}
\end{equation}
as the difference in the binding energy of the anharmonic bipolaron from the 
harmonic bipolaron.  This reduction in bipolaron binding energy is plotted
in Figures~2(a) and (b) as a function of the harmonic bipolaron binding
energy (for fixed values of $\alpha_{anh}$) and as a function of $\alpha_{anh}$
(for fixed values of $U$), respectively.  Notice how even a small value of
the anharmonicity produces a large reduction of the bipolaron binding
energy, and that the reduction increases as the electron-phonon interaction
strength increases (the phonon frequency is fixed at $\Omega/t^*=0.5$).

\section{Mean-Field-Theory Analysis}

We consider the case of hopping between nearest neighbors on a hypercubic
lattice in $d$-dimensions.  The only nonzero matrix elements are then 
$t_{ij}=t^*/2\sqrt{d}$ when $i$ and $j$ are nearest neighbors.  As described
above, $t^*$ is chosen to be the energy unit.  This choice of scaling the
hopping matrix elements inversely as the square root of the dimensionality
is made so that the theory has a nontrivial limit\cite{metzner_vollhardt}
as $d\rightarrow\infty$.

We employ a mean-field-theory analysis to the pseudospin form of the 
effective Hamiltonian.  Two types of phase transitions occur: (1) staggered
order along the $z$-axis [corresponding to CDW order at the 
``antiferromagnetic'' ($\pi,\pi,\ldots$) point]; and (2) staggered order along 
the $x$-axis (corresponding to SC order with a zero-momentum pair-field state). 
The mean-field theory becomes exact in high dimensions, and should provide an
upper bound to the transition temperatures in finite dimensions (because
nonlocal quantum fluctuations should reduce the transition temperature further).

The mean-field theory is constructed by determining the molecular field
at each lattice site and equating the expectation value of the magnetization
with that of a free spin in an external magnetic field equal to the molecular
field, ${\bf h}_{mol.}$, yielding
\begin{equation}
\langle {\bf J}\rangle = \frac{{\bf h}_{mol.}}{|{\bf h}_{mol.}|}
\frac{1}{2} \tanh \frac{1}{2}\beta |{\bf h}_{mol.}|\quad,
\label{eq: hmol}
\end{equation}
as first described by Gorter\cite{gorter}.  The difficult part of the
calculation involves a correct determination of the molecular field.  One
must be certain to properly count the contributions from each of the diagrams
in Figure~1.  Note that the total number of distinct
nearest-neighbor pairs corresponding to Figures~1(a) and 1(b) is $Nd$, with
$N$ the number of lattice sites, and
each pair appears twice in the unrestricted summations.  There are two classes
of second-neighbor diagrams corresponding to Figure~1(c): those where $j$ and 
$k$ are not parallel [of the $(1,1,0,\ldots)$ form] and those where $j$ and $k$
are parallel [of the $(2,0,0,\ldots)$ form]. In the first case, there are
$Nd(d-1)$ pairs, with each pair appearing four times in the unrestricted 
summation.  In the second case, there are $Nd$ pairs, with each pair 
appearing twice in the unrestricted summation.  Finally there are
$\frac{1}{2}Nd(d-1)$ distinct squares corresponding to Figure~1(d) with
each square appearing eight times in the unrestricted summation.  Using these
results, it is a straightforward exercise to rewrite the Hamiltonian as
a summation over distinct pairs (and squares) and then extract the molecular
fields for the corresponding ordered phases.

A hypercubic lattice is bipartite, so it divides into two sublattices $A$ and
$B$, where the nearest-neighbor hopping occurs only from sublattice $A$ to
sublattice $B$ or vice versa.  The paramagnetic high-temperature phase 
corresponds to a uniform magnetization of the pseudospins on each sublattice:
\begin{equation}
\langle {\bf J}_A\rangle =\langle {\bf J}_B\rangle =:\frac{1}{2}
 m {\bf e}_z = \frac{1}{2}(\rho _e -1){\bf e}_z\quad,
\label{eq: paramag}
\end{equation}
where ${\bf e}_z$ is the unit vector along the $z$-axis and $\rho_e$ is the
electron concentration.  The self-consistent equation for the pseudospin
magnetization becomes
\begin{eqnarray}
m&=&\tanh \frac{1}{2}\beta \biggl \{2(\mu -U)+md\left [
-j_{\parallel}^{(2)}-j_{\parallel}^{(4)} +(2d-1)(j_{\parallel}^{\prime}-
j_{\parallel}^{\prime\prime})+(d-1)(\delta+\frac{1}{2}\nu)\right ] \cr
&\quad&\quad\quad\quad\quad\quad\quad\quad\quad\quad
+m^3d(d-1)[2\alpha-\delta-\frac{1}{2}\nu ] \biggr \}
\quad.
\label{eq: selfcons}
\end{eqnarray}
The dependence of the chemical potential $\mu$ upon the electron concentration
$\rho_e$ can easily be determined by inverting Eq.~(\ref{eq: selfcons}).

The transition temperature to the
commensurate charge-density-wave phase occurs at a temperature where
the pseudospin magnetization satisfies
\begin{equation}
\langle {\bf J}_A\rangle =\frac{1}{2}(m+m^{\prime}) {\bf e}_z \quad,\quad
\langle {\bf J}_B\rangle =\frac{1}{2}(m-m^{\prime}) {\bf e}_z \quad,
\label{eq: cdwmag}
\end{equation}
in the limit $m^{\prime}\rightarrow 0$.  We only consider the transition to
a commensurate CDW, because transitions to incommensurate phases can only
occur if the frustration induced by the fourth-order terms in the effective 
Hamiltonian becomes large enough.  Since the validity of the truncated 
strong-coupling expansion fails when the fourth-order terms are comparable
in size to the second-order terms, we ignore the complication of incommensurate
order here.

The transition temperature is then easily found to be
\begin{eqnarray}
T_c&=&\frac{1}{2}\rho_e(2-\rho_e)\{dj_{\parallel}^{(2)}-d^2[
6j_{\parallel}^{\prime}+2j_{\parallel}^{\prime\prime}-\delta+\frac{3}{2}\nu
+(2\alpha-\delta-\frac{1}{2}\nu)(1-\rho_e)^2]\cr
&\quad&\quad\quad\quad+d[j_{\parallel}^{(4)}+3j_{\parallel}^{\prime}+
j_{\parallel}^{\prime\prime}-\delta+\frac{3}{2}\nu
+(2\alpha-\delta-\frac{1}{2}\nu)(1-\rho_e)^2]\}\quad .
\label{eq: tc_cdw}
\end{eqnarray}
Note that this expression differs slightly (in the coefficient of the $\delta$
term) from that given previously\cite{freericks_strong}, and corrects a 
typo in that work.  Explicit formulae for the parameters appearing in
Eq.~(\ref{eq: tc_cdw}) appear in the Appendix.

Likewise, the superconducting transition temperature is determined by finding 
the temperature where the pseudospin magnetization satisfies
\begin{equation}
\langle J_A^z\rangle =\langle J_B^z\rangle = \frac{1}{2}m\quad ,\quad
\langle J_A^x\rangle = -\langle J_B^x\rangle = \frac{1}{2}m^{\prime}\quad ,
\label{eq: scmag}
\end{equation}
in the limit $m^{\prime}\rightarrow 0$. The transition temperature is
\begin{eqnarray}
T_c&=&\frac{\rho_e-1}{\ln [\rho_e/(2-\rho_e)]}\{ dj_{\perp}^{(2)}+
d^2[ 4j_{\perp}^{\prime}-2j_{\perp}^{\prime\prime}+
2\beta -\gamma +2\epsilon -\mu +(2\beta - \gamma -2\epsilon +\mu)(1-\rho_e)^2
]\cr
&\quad&\quad\quad\quad\quad+d[j_{\perp}^{(4)}-2j_{\perp}^{\prime}+
j_{\perp}^{\prime\prime}-2\beta+\gamma-2\epsilon+\mu
+(-2\beta+\gamma+2\epsilon-\mu)(1-\rho_e)^2] \}\quad.
\label{eq: tc_sc}
\end{eqnarray}
Explicit formulae for the parameters appearing in Eq.~(\ref{eq: tc_sc}) 
appear in the Appendix.

The above expressions for the transition temperatures to CDW or SC order
are evaluated below in the infinite-dimensional limit.  In this case, there
is no contribution from the fourth-order terms that are multiplied by a
linear power in $d$, because they scale like $t^{*4}/d\rightarrow 0$ in the
large-dimensional limit.

At half-filling $(\rho_e=1)$, the CDW-phase is expected to be the ground
state if either the anharmonicity or the Coulomb repulsion is not too
large.  Figure~3 plots the CDW transition temperature at half filling
for different values of the anharmonicity 3(a) and the Coulomb repulsion 3(b).
Both the second-order approximations (which monotonically diverge as the
interaction strength approaches zero) and the fourth-order approximations
(which properly show a peak in $T_c$ as a function of the coupling) are
plotted.  In Figure~3(a), four different values of the anharmonicity are
shown: $\alpha_{anh}=0.0$ (the harmonic case) (solid line); $\alpha_{anh}=0.001$
(dotted line); $\alpha_{anh}=0.005$ (dashed line); and $\alpha_{anh}=0.015$
(dash-dotted line).  It is apparent that even though the second-order
approximation shows large enhancements to the transition temperature as
the anharmonicity is increased, the fourth-order approximation indicates
that the maximal CDW transition temperature actually decreases as the
anharmonicity increases.  Furthermore, the value of coupling strength where
the maximum occurs increases as a function of anharmonicity.  This is
to be expected since the anharmonicity acts in some respects like a retarded
Coulomb repulsion.  What is surprising is that relatively small values of
anharmonicity have such large effects on the transition temperature.

In Figure~3(b), the CDW transition temperature at half filling is plotted
for four different values of $U_c(\alpha_{anh})$.  The same values of 
$\alpha_{anh}$ are used as in Figure~3(a).  Note that once again the maximal
$T_c$ decreases as $\alpha_{anh}$ increases (implying that $U_c$ increases).
Furthermore, the peak does not move as rapidly to larger values of the
coupling strength, indicating that the retardation effects are rather
strong even at the relatively large phonon frequency of $\Omega/t^*=0.5$.

Note that these strong-coupling phase diagrams, may be more accurate than those
of the harmonic model, because it is known that in the case of either
an anharmonic interaction, or a Coulomb repulsion, that the CDW instability
only occurs when the coupling strength $U$ is large enough in magnitude.
Hence the true $T_c$ does vanish at a finite value of $U$, as it does in
the fourth-order approximation.

Our analysis at half filling ignored the possibility of SC order.  In fact, the
system will go superconducting at half filling in the regime where the 
CDW transition temperature has been suppressed to zero, but since this regime
corresponds to a region where the fourth-order approximations are breaking
down, we have not complicated Figure~3 by including the superconducting
solutions at small $|U|$.  Rather, we examine what happens
as the system is doped away from half-filling.  In Figure~4, the phase diagram
is plotted for the coupling strength $U=-1.5625t^*$ which lies at the peak
of the CDW transition temperature curve when $\alpha_{anh}=0$.  This case
represents a lower limit of applicability of the strong-coupling expansion.
In Figure~4(a), the anharmonicity varies from $\alpha_{anh}=$ 0, 0.003, 0.005,
0.01.  The solid lines denote CDW solutions, and the dotted lines are SC.
Note that the CDW-SC phase boundary lies at exponentially small densities
in the harmonic case.  As the anharmonicity is turned on, the CDW transition
temperature is suppressed, and the SC transition temperature is enhanced, so
the CDW-SC phase boundary moves in towards half filling.  When 
$\alpha_{anh}=0.005$, the CDW phase has disappeared.  The enhancement of the 
SC transition temperature continues as $\alpha_{anh}$ increases, until it is 
significantly larger than the maximal CDW transition temperature of the
harmonic Holstein model.  Hence, the
strong-coupling approximation, through fourth order, predicts a large 
enhancement in the SC transition temperature relative to the harmonic case.
It is not clear whether this result is an artifact of the approximation,
or is a real effect.  Note further that the electron density where the maximal
superconducting $T_c$ occurs is near the band edges.  This is similar in
spirit to Hirsch's mechanism, but, as far as we can tell is unrelated, because
in the strong-coupling limit the effect occurs both at the top and the bottom
of the band, since the approximation explicitly retains electron-hole symmetry.

In Figure~4(b), the same phase diagrams are plotted, this time using $U_c(
\alpha_{anh})$ with the same values of $\alpha_{anh}$ as in Figure~4(a).
Clearly one can see that the effect of the Coulomb repulsion is quite different
from the anharmonicity.  The phase diagram does not change much, and the
large enhancement of $T_c$ in the SC channel does not occur.

In order to check to see whether these results are generic, or occur simply
because one is close to the limiting region where the approximations are
expected to hold, we have also calculated the phase diagrams for a stronger
value of the interaction strength $U=-4.0t^*$.  In Figure~5(a), four 
different values of $\alpha_{anh}$ (0.0, 0.01, 0.03, 0.05) are plotted.
Here the results are similar to those found in Figure~4, except the 
anharmonicity initially causes the CDW $T_c$ to rise, because of the reduction
in the bipolaron binding energy.  The SC transition temperature still has
a large maximum at low densities, and is significantly enhanced relative to
the harmonic case.  Furthermore, the CDW-SC phase boundary continues to move
toward half-filling until it disappears at $\alpha_{anh}\approx 0.03$.  In
Figure~5(b), we plot the same phase diagrams, this time using 
$U_c(\alpha_{anh}$).  Once again, the phase diagram displays very different 
behavior, with the CDW-SC phase transition remaining at exponentially
small densities.

Finally, we study how the critical electron density, where the CDW-SC phase
boundary lies, evolves as functions of $U$ and $\alpha_{anh}$.  In the
anharmonic Holstein model, we see in Figure~6(a) that the phase boundary
moves very rapidly as the anharmonicity is turned on.  This indicates how
the anharmonicity strongly favors SC solutions relative to CDW order.
In Figure~6(b), we show the analogous plots of the critical electron 
density for the harmonic Holstein-Hubbard model, with $U_c$ chosen from
Eq.~(\ref{eq: ucalpha}), and the same values of $\alpha_{anh}$.  Clearly the
anharmonic behavior is not easily mimicked by an instantaneous Coulomb
repulsion, and the retardation effects cannot be neglected.

\section{Conclusions}

The strong-coupling expansion for the anharmonic Holstein-Hubbard model has
been presented through fourth order in the hopping.  This result extends the 
analysis of the harmonic case\cite{freericks_strong}.  We find some interesting
results from this analysis.  First, the anharmonicity reduces the bipolaron 
binding energy and second, it reduces the
the equilibrium phonon-coordinate spacing between one electron
at a site and either zero or two electrons at a site.  The first effect is
expected to cause an enhancement to transition temperatures in the 
strong-coupling regime, and the second should enhance the Franck-Condon
overlap factors for superconducting order, favoring the SC phase relative
to the CDW. Similarly, a Coulomb repulsion will reduce the bipolaron binding
energy, but does not alter the Franck-Condon overlaps.  We find that in the
CDW phase, the maximal transition temperature actually decreases when either
anharmonicity or Coulomb repulsion are turned on.  For the superconducting
phase, the enhancement of the Franck-Condon overlap factors (equivalent
to a widening of the polaron band) causes a large enhancement of the 
SC transition temperature at low electron density, even for moderate values of
the anharmonicity.  In no case do we find that the effect of the anharmonicity
is easily mimicked by an effective Coulomb repulsion.

Since the strong-coupling expansion is expected to fail in both the 
low-electron-concentration regime, and when the effective bipolaron binding
energy is no longer much larger than the hopping integral, it is possible that
this large enhancement of the superconducting transition temperature is just
an artifact of the current approximation.  It is important to compare these
approximations to exact quantum Monte Carlo simulations of the transition
temperatures of the anharmonic Holstein model.  Work in this direction is
currently in progress.

\acknowledgments
We would like to acknowledge useful conversations with J. Hirsch and M. Jarrell.
J.~K.~F. acknowledges the Donors of The Petroleum Research Fund, administered 
by the American Chemical Society, for partial support of this research 
(ACS-PRF\# 29623-GB6) and an Oak Ridge Associated University Junior Faculty 
Enhancement Award for partial support of this research. 
G.~D.~M. acknowledges support by the University of Tennessee, 
and by Oak Ridge National Laboratory, managed by 
Lockhead Martin Energy Research Corporation for the
U.S. Department of Energy under contract number
DE-AC05-96OR22464.

\appendix

\section{Appendix: Parameters of the fourth-order
effective pseudospin Hamiltonian}

In this appendix explicit expressions are given
for the parameters that appear in the
fourth-order effective pseudospin Hamiltonian, summarized in
Eqs.~(\ref{eq: h4bfinal}), (\ref{eq: h4cfinal}),
and (\ref{eq: h4dfinal}).  The notation used is that given in the text.

First the parameters in Eq.~(\ref{eq: h4bfinal}):
\begin{eqnarray}
j_{\parallel}^{(4)}&=&-8t_{ij}^4 \biggl \{  \sum_{l,l^{\prime},m,
m^{\prime},n,n^{\prime}=0 \atop m+m^{\prime}\ne 0}^{\infty} \cr
&\biggl [&
\frac{\langle +0|n\rangle\langle n|+m\rangle\langle +m|l\rangle\langle l|+0
\rangle\langle -0|n^{\prime}\rangle
\langle n^{\prime}|-m^{\prime}\rangle\langle -m^{\prime}|l^{\prime}\rangle
\langle l^{\prime}|-0\rangle}
{[E_+(0)+E_-(0)-E(n)-E(n^{\prime})][E_+(0)+E_-(0)-E_+(m)-E_-(m^{\prime})]
}\cr 
&\quad&\quad\quad\quad\quad\quad\times\frac{1}{E_+(0)+E_-(0)-E(l)-E(l^{\prime})}
\cr
&+&
\frac{\langle +0|n\rangle\langle n|-m\rangle\langle -m|l\rangle\langle l|+0
\rangle\langle -0|n^{\prime}\rangle
\langle n^{\prime}|+m^{\prime}\rangle\langle +m^{\prime}|l^{\prime}\rangle
\langle l^{\prime}|-0\rangle}
{[E_+(0)+E_-(0)-E(n)-E(n^{\prime})][E_+(0)+E_-(0)-E_+(m)-E_-(m^{\prime})]} \cr
&\quad&\quad\quad\quad\quad\quad\times\frac{1}{E_+(0)+E_-(0)-E(l)-E(l^{\prime})}
\biggr ]\cr
&\quad&
-\sum_{l,l^{\prime},n,n^{\prime}=0}^{\infty} \biggl [
\frac{\langle +0|n\rangle\langle n|+0\rangle\langle -0|n^{\prime}\rangle
\langle n^{\prime}|-0\rangle\langle +0|l\rangle\langle l|+0\rangle
\langle -0|l^{\prime}\rangle\langle l^{\prime}|-0\rangle}
{[E_+(0)+E_-(0)-E(n)-E(n^{\prime})]^2[E_+(0)+E_-(0)-E(l)-E(l^{\prime})]}\cr 
&\quad&\quad\quad\quad+
\frac{\langle +0|n\rangle\langle n|-0\rangle\langle -0|n^{\prime}\rangle
\langle n^{\prime}|+0\rangle\langle -0|l\rangle\langle l|+0\rangle
\langle +0|l^{\prime}\rangle\langle l^{\prime}|-0\rangle}
{[E_+(0)+E_-(0)-E(n)-E(n^{\prime})]^2[E_+(0)+E_-(0)-E(l)-E(l^{\prime})]}
 \biggr ] \biggr \} \quad ,
\label{eq: j_par_4_c}
\end{eqnarray}
\begin{eqnarray}
j_{\perp}^{(4)}&=&-8t_{ij}^4
\biggl \{  \sum_{l,l^{\prime},m,
m^{\prime},n,n^{\prime}=0 \atop m+m^{\prime}\ne 0}^{\infty}\cr
& \biggl [&
\frac{\langle +0|n\rangle\langle n|+m\rangle\langle +m|l\rangle\langle l|-0
\rangle\langle -0|n^{\prime}\rangle
\langle n^{\prime}|-m^{\prime}\rangle\langle -m^{\prime}|l^{\prime}\rangle
\langle l^{\prime}|+0\rangle}
{[E_+(0)+E_-(0)-E(n)-E(n^{\prime})][E_+(0)+E_-(0)-E_+(m)-E_-(m^{\prime})]
}\cr 
&\quad&\quad\quad\quad\quad\quad\times\frac{1}{E_+(0)+E_-(0)-E(l)-E(l^{\prime})}
\cr
&+&
\frac{\langle +0|n\rangle\langle n|-m\rangle\langle -m|l\rangle\langle l|-0
\rangle\langle -0|n^{\prime}\rangle
\langle n^{\prime}|+m^{\prime}\rangle\langle +m^{\prime}|l^{\prime}\rangle
\langle l^{\prime}|+0\rangle}
{[E_+(0)+E_-(0)-E(n)-E(n^{\prime})][E_+(0)+E_-(0)-E_+(m)-E_-(m^{\prime})]} \cr
&\quad&\quad\quad\quad\quad\quad\times\frac{1}{E_+(0)+E_-(0)-E(l)-E(l^{\prime})}
\biggr ]\cr
&\quad& -\sum_{l,l^{\prime},n,n^{\prime}=0}^{\infty} \biggl [
\frac{\langle +0|n\rangle\langle n|+0\rangle\langle -0|n^{\prime}\rangle
\langle n^{\prime}|-0\rangle\langle +0|l\rangle\langle l|-0\rangle
\langle -0|l^{\prime}\rangle\langle l^{\prime}|+0\rangle}
{[E_+(0)+E_-(0)-E(n)-E(n^{\prime})]^2[E_+(0)+E_-(0)-E(l)-E(l^{\prime})]}\cr 
&\quad&\quad\quad\quad+
\frac{\langle +0|n\rangle\langle n|-0\rangle\langle -0|n^{\prime}\rangle
\langle n^{\prime}|+0\rangle\langle -0|l\rangle\langle l|-0\rangle
\langle +0|l^{\prime}\rangle\langle l^{\prime}|+0\rangle}
{[E_+(0)+E_-(0)-E(n)-E(n^{\prime})]^2[E_+(0)+E_-(0)-E(l)-E(l^{\prime})]}
 \biggr ] \biggr \} \quad .
\label{eq: j_per_4_c}
\end{eqnarray}

Next the parameters in Eq.~(\ref{eq: h4cfinal}):
\begin{eqnarray}
&\quad&j_{\parallel}^{\prime}=4t_{ij}^2t_{jk}^2 \biggl \{  
\sum_{l,l^{\prime},m,n,n^{\prime}=0}^{\infty} \cr
&\quad&\quad\quad\quad\quad
\frac{\langle +0|n\rangle\langle n|+0\rangle\langle -0|n^{\prime}\rangle
\langle n^{\prime}|+m\rangle\langle +m|l^{\prime}\rangle\langle l^{\prime}|-0
\rangle\langle +0|l\rangle\langle l|+0\rangle}
{[E_+(0)+E_-(0)-E(n)-E(n^{\prime})][2E_+(0)+E_-(0)-E_+(m)-E(n)-E(l)]}\cr 
&\quad&\quad\quad\quad\quad\quad\quad\quad\quad\times\left [
\frac{1}{E_+(0)+E_-(0)-E(l)-E(l^{\prime})}+
\frac{1}{E_+(0)+E_-(0)-E(n)-E(l^{\prime})}\right ]\cr
&\quad&+2\sum_{l,l^{\prime},n,n^{\prime}=0 \atop m\ne 0}^{\infty} 
\frac{\langle +0|n\rangle\langle n|+0\rangle\langle -0|n^{\prime}\rangle
\langle n^{\prime}|-m\rangle\langle -m|l^{\prime}\rangle\langle l^{\prime}|-0
\rangle\langle +0|l\rangle\langle l|+0\rangle}
{[E_+(0)+E_-(0)-E(n)-E(n^{\prime})][E_+(0)-E_+(m)]
[E_+(0)+E_-(0)-E(l)-E(l^{\prime})]}\cr&\quad&
-2\sum_{l,l^{\prime},n,n^{\prime}=0}^{\infty}
\frac{\langle +0|n\rangle\langle n|+0\rangle\langle -0|n^{\prime}\rangle
\langle n^{\prime}|-0\rangle\langle -0|l^{\prime}\rangle\langle l^{\prime}|-0
\rangle\langle +0|l\rangle\langle l|+0\rangle}
{[E_+(0)+E_-(0)-E(n)-E(n^{\prime})]^2[E_+(0)+E_-(0)-E(l)-E(l^{\prime})]}
\biggr \} \quad ,
\label{eq: j_par_4_a}
\end{eqnarray}
\begin{eqnarray}
j_{\parallel}^{\prime\prime}&=&
-4t_{ij}^2t_{jk}^2\sum_{l,l^{\prime},m,n,n^{\prime} =0}^{\infty}
\frac{\langle +0|n\rangle\langle n|+m\rangle\langle +m|l^{\prime}
\rangle\langle l^{\prime}|+0\rangle\langle -0|n^{\prime}\rangle\langle
n^{\prime}|-0\rangle\langle +0|l\rangle\langle l|+0\rangle}
{E_+(0)+E_-(0)-E(n)-E(n^{\prime})}
\cr
&\quad&\quad\quad\quad\quad\times\frac{1}
{[2E_+(0)+E_-(0)-E_+(m)-E(l)-E(n^{\prime})]
[E_+(0)+E_-(0)-E(l^{\prime})-E(n^{\prime})]}\quad ,
\label{eq: j_par_4_b}
\end{eqnarray}
\begin{eqnarray}
&\quad&j_{\perp}^{\prime}=-4t_{ij}^2t_{jk}^2\biggl \{
\sum_{l,l^{\prime},m,n,n^{\prime}=0}^{\infty} \cr
&\quad&\quad\quad\frac{\langle +0|n\rangle\langle n|+0\rangle
\langle -0|n^{\prime}\rangle\langle n^{\prime}|+m\rangle\langle +m|l^{\prime}
\rangle\langle l^{\prime}|+0\rangle\langle +0|l\rangle\langle l|-0\rangle}
{[E_+(0)+E_-(0)-E(n)-E(n^{\prime})][2E_+(0)+E_-(0)-E_+(m)-E(n)-E(l)]}\cr
&\quad&\quad\quad\quad\quad\quad\times\left [ \frac{1}{E_+(0)+E_-(0)-E(l)-
E(l^{\prime})}+\frac{1}{E_+(0)+E_-(0)-E(n)-E(l^{\prime})}\right ]\cr
&\quad&\quad +2\sum_{l,l^{\prime},n,n^{\prime}=0 \atop m\ne 0}^{\infty}
\frac{\langle +0|n\rangle\langle n|+0\rangle \langle -0|n^{\prime}\rangle
\langle n^{\prime}|-m\rangle \langle -m|l^{\prime}\rangle \langle l^{\prime}|
+0\rangle \langle +0|l\rangle\langle l|-0\rangle}
{[E_+(0)+E_-(0)-E(n)-E(n^{\prime})][E_+(0)-E_+(m)][E_+(0)+E_-(0)-E(l)-E(
l^{\prime})]}\cr
&\quad&\quad -\sum_{l,l^{\prime},n,n^{\prime}=0}^{\infty}
\frac{\langle +0|n\rangle\langle n|+0\rangle \langle -0|n^{\prime}\rangle
\langle n^{\prime}|-0\rangle \langle -0|l^{\prime}\rangle \langle l^{\prime}|
+0\rangle \langle +0|l\rangle\langle l|-0\rangle}
{[E_+(0)+E_-(0)-E(n)-E(n^{\prime})][E_+(0)+E_-(0)-E(l)-E(l^{\prime})]}\cr
&\quad&\quad\quad\quad\quad\quad\times\left [ \frac{1}{E_+(0)+E_-(0)-E(n)-
E(n^{\prime})}+\frac{1}{E_+(0)+E_-(0)-E(l)-E(l^{\prime})}\right ]
\biggr \}\quad ,
\label{eq: j_per_4a}
\end{eqnarray}
\begin{eqnarray}
&\quad&j_{\perp}^{\prime\prime}=4t_{ij}^2t_{jk}^2\biggl \{
\sum_{l,l^{\prime},m,n,n^{\prime}=0}^{\infty} \cr
&\quad&\quad\quad\quad\quad\quad\frac{\langle +0|n\rangle\langle n|+m\rangle
\langle +m|l^{\prime}\rangle\langle l^{\prime}|+0\rangle\langle -0|n^{\prime}
\rangle\langle n^{\prime}|+0\rangle\langle +0|l\rangle\langle l|-0\rangle}
{[E_+(0)+E_-(0)-E(n)-E(n^{\prime})][2E_+(0)+E_-(0)-E_+(m)-E(n^{\prime})-E(l)]}
\cr
&\quad&\quad\quad\quad\quad\quad\quad\quad\quad
\times \frac{1}{E_+(0)+E_-(0)-E(l)-E(l^{\prime})}\cr
&\quad&\quad +2\sum_{l,l^{\prime},n,n^{\prime}=0 \atop m\ne 0}^{\infty}
\frac{\langle +0|n\rangle\langle n|-m\rangle \langle -m|l^{\prime}\rangle
\langle l^{\prime}|+0\rangle \langle -0|n^{\prime}\rangle \langle n^{\prime}|
+0\rangle \langle +0|l\rangle\langle l|-0\rangle}
{[E_+(0)+E_-(0)-E(n)-E(n^{\prime})][E_+(0)-E_+(m)][E_+(0)+E_-(0)-E(l)-E(
l^{\prime})]}\cr
&\quad&\quad -2\sum_{l,l^{\prime},n,n^{\prime}=0}^{\infty}
\frac{\langle +0|n\rangle\langle n|-0\rangle \langle -0|n^{\prime}\rangle
\langle n^{\prime}|+0\rangle \langle -0|l^{\prime}\rangle \langle l^{\prime}|
+0\rangle \langle +0|l\rangle\langle l|-0\rangle}
{[E_+(0)+E_-(0)-E(n)-E(n^{\prime})]^2[E_+(0)+E_-(0)-E(l)-E(l^{\prime})]}
\biggr \}\quad .
\label{eq: j_per_4b}
\end{eqnarray}

Finally the parameters in Eq.~(\ref{eq: h4dfinal}):
\begin{eqnarray}
\alpha&=&4t_{ij}t_{jk}t_{kl}t_{li}\sum_{l,l^{\prime},n,n^{\prime}=0}^{\infty}
\frac{\langle +0|n\rangle\langle n|+0\rangle\langle -0|n^{\prime}\rangle
\langle n^{\prime}|-0\rangle\langle +0|l\rangle\langle l|+0\rangle
\langle +0|l^{\prime}\rangle\langle l^{\prime}|+0\rangle}
{[E_+(0)+E_-(0)-E(n)-E(n^{\prime})][E_+(0)+E_-(0)-E(l)-E(n^{\prime})]}\cr
&\quad&\quad\quad\quad\quad\quad\times\frac{1}
{E_+(0)+E_-(0)-E(l^{\prime})-E(n^{\prime})}\quad ,
\label{eq: alpha}
\end{eqnarray}
\begin{eqnarray}
\beta&=&-2t_{ij}t_{jk}t_{kl}t_{li}\sum_{l,l^{\prime},n,n^{\prime}=0}^{\infty}
\biggl \{ \frac{\langle +0|n\rangle\langle n|-0\rangle\langle -0|n^{\prime}
\rangle\langle n^{\prime}|+0\rangle\langle +0|l\rangle\langle l|+0\rangle
\langle +0|l^{\prime}\rangle\langle l^{\prime}|+0\rangle}
{[E_+(0)+E_-(0)-E(n)-E(n^{\prime})][E_+(0)+E_-(0)-E(l)-E(n)]}\cr
&\quad&\quad\quad\quad\quad\quad\quad\quad\quad\quad
\times\frac{1}{E_+(0)+E_-(0)-E(l^{\prime})-E(n)}\cr
&\quad&\quad\quad\quad\quad\quad
+\frac{\langle +0|n\rangle\langle n|+0\rangle\langle -0|
n^{\prime}\rangle\langle n^{\prime}|+0\rangle\langle +0|l\rangle
\langle l|+0\rangle\langle +0|l^{\prime}\rangle\langle l^{\prime}|-0\rangle}
{E_+(0)+E_-(0)-E(n)-E(n^{\prime})}\cr
&\quad&\quad\quad\quad\quad\quad
\times \biggl [ \frac{1}{[E_+(0)+E_-(0)-E(l^{\prime})-E(n)]
[E_+(0)+E_-(0)-E(l)-E(l^{\prime})]}\cr
&\quad&\quad\quad\quad\quad\quad\quad +
\frac{1}{[E_+(0)+E_-(0)-E(l)-E(n^{\prime})][E_+(0)+E_-(0)-E(l)-E(l^{\prime})]}
\cr
&\quad&\quad\quad\quad\quad\quad\quad +
\frac{1}{[E_+(0)+E_-(0)-E(l)-E(n^{\prime})][E_+(0)+E_-(0)-E(l^{\prime})-
E(n^{\prime})]}\biggr ]\biggr \}\quad ,
\label{eq: beta}
\end{eqnarray}
\begin{eqnarray}
\gamma&=&4t_{ij}t_{jk}t_{kl}t_{li}\sum_{l,l^{\prime},n,n^{\prime}=0}^{\infty}
\biggl \{ \frac{\langle +0|n\rangle\langle n|+0\rangle\langle -0|n^{\prime}
\rangle\langle n^{\prime}|+0\rangle\langle +0|l\rangle\langle l|-0\rangle
\langle +0|l^{\prime}\rangle\langle l^{\prime}|+0\rangle}
{[E_+(0)+E_-(0)-E(n)-E(n^{\prime})][E_+(0)+E_-(0)-E(l)-E(n^{\prime})]}\cr
&\quad&\quad\quad\quad\quad\quad\quad\quad\quad\quad
\times\frac{1}{E_+(0)+E_-(0)-E(l)-E(l^{\prime})}\cr
&\quad&\quad\quad\quad\quad\quad
+\frac{\langle +0|n\rangle\langle n|+0\rangle\langle -0|
n^{\prime}\rangle\langle n^{\prime}|+0\rangle\langle +0|l\rangle
\langle l|+0\rangle\langle +0|l^{\prime}\rangle\langle l^{\prime}|-0\rangle}
{[E_+(0)+E_-(0)-E(n)-E(n^{\prime})][E_+(0)+E_-(0)-E(l)-E(n)]}\cr
&\quad&\quad\quad\quad\quad\quad
\times \left [ \frac{1}{E_+(0)+E_-(0)-E(l^{\prime})-E(n)}
+\frac{1}{E_+(0)+E_-(0)-E(l)-E(l^{\prime})}
\right ]\biggr \}\quad ,
\label{eq: gamma}
\end{eqnarray}
\begin{eqnarray}
\delta&=&-4t_{ij}t_{jk}t_{kl}t_{li}\sum_{l,l^{\prime},n,n^{\prime}=0}^{\infty}
\biggl \{ \frac{\langle +0|n\rangle\langle n|+0\rangle\langle -0|n^{\prime}
\rangle\langle n^{\prime}|-0\rangle\langle +0|l\rangle\langle l|+0\rangle
\langle -0|l^{\prime}\rangle\langle l^{\prime}|-0\rangle}
{[E_+(0)+E_-(0)-E(n)-E(n^{\prime})][E_+(0)+E_-(0)-E(l)-E(n^{\prime})]}\cr
&\quad&\quad\quad\quad\quad\quad\quad\quad\quad\quad
\times\frac{1}{E_+(0)+E_-(0)-E(l)-E(l^{\prime})}\cr
&\quad&\quad\quad\quad\quad\quad+
\frac{\langle +0|n\rangle\langle n|+0\rangle\langle -0|n^{\prime}
\rangle\langle n^{\prime}|-0\rangle\langle -0|l\rangle\langle l|-0\rangle
\langle +0|l^{\prime}\rangle\langle l^{\prime}|+0\rangle}
{[E_+(0)+E_-(0)-E(n)-E(n^{\prime})][E_+(0)+E_-(0)-E(l)-E(n)]}\cr
&\quad&\quad\quad\quad\quad\quad\quad\quad\quad\quad
\times\frac{1}{E_+(0)+E_-(0)-E(l)-E(l^{\prime})}\biggr \}\quad ,
\label{eq: delta}
\end{eqnarray}
\begin{eqnarray}
\epsilon&=&2t_{ij}t_{jk}t_{kl}t_{li}\sum_{l,l^{\prime},n,n^{\prime}=0}^{\infty}
\biggl \{ 2\frac{\langle +0|n\rangle\langle n|-0\rangle\langle -0|n^{\prime}
\rangle\langle n^{\prime}|+0\rangle\langle +0|l\rangle\langle l|+0\rangle
\langle -0|l^{\prime}\rangle\langle l^{\prime}|-0\rangle}
{E_+(0)+E_-(0)-E(n)-E(n^{\prime}}\cr
&\quad&\quad\quad\quad\quad\quad\quad\quad\quad\quad\quad
\times\frac{1}{2E_+(0)+2E_-(0)-E(n)-E(n^{\prime})-E(l)-E(l^{\prime})}\cr
&\quad&\quad\quad\quad\quad\quad\quad
\times\left [\frac{1}{E_+(0)+E_-(0)-E(l^{\prime})-E(n^{\prime})}
+\frac{1}{E_+(0)+E_-(0)-E(l)-E(n)}\right ]\cr
&\quad&\quad\quad\quad\quad\quad+
\frac{\langle +0|n\rangle\langle n|+0\rangle\langle -0|n^{\prime}
\rangle\langle n^{\prime}|-0\rangle\langle +0|l\rangle\langle l|-0\rangle
\langle -0|l^{\prime}\rangle\langle l^{\prime}|+0\rangle}
{[E_+(0)+E_-(0)-E(n)-E(n^{\prime})][E_+(0)+E_-(0)-E(l)-E(n^{\prime})]}\cr
&\quad&\quad\quad\quad\quad\quad\quad
\times\left [\frac{1}{E_+(0)+E_-(0)-E(l^{\prime})-E(n^{\prime})}
+\frac{1}{E_+(0)+E_-(0)-E(l)-E(l^{\prime})}\right ]\cr
&\quad&\quad\quad\quad\quad\quad+
\frac{\langle +0|n\rangle\langle n|+0\rangle\langle -0|n^{\prime}
\rangle\langle n^{\prime}|-0\rangle\langle -0|l\rangle\langle l|+0\rangle
\langle +0|l^{\prime}\rangle\langle l^{\prime}|-0\rangle}
{[E_+(0)+E_-(0)-E(n)-E(n^{\prime})][E_+(0)+E_-(0)-E(l)-E(n)]}\cr
&\quad&\quad\quad\quad\quad\quad\quad
\times\left [\frac{1}{E_+(0)+E_-(0)-E(l^{\prime})-E(n)}
+\frac{1}{E_+(0)+E_-(0)-E(l)-E(l^{\prime})}\right ]\biggr \}\quad ,
\label{eq: epsilon}
\end{eqnarray}
\begin{eqnarray}
\mu&=&-2t_{ij}t_{jk}t_{kl}t_{li}\sum_{l,l^{\prime},n,n^{\prime}=0}^{\infty}
\biggl \{ 2\frac{\langle +0|n\rangle\langle n|-0\rangle\langle -0|n^{\prime}
\rangle\langle n^{\prime}|-0\rangle\langle +0|l\rangle\langle l|+0\rangle
\langle -0|l^{\prime}\rangle\langle l^{\prime}|+0\rangle}
{E_+(0)+E_-(0)-E(n)-E(n^{\prime}}\cr
&\quad&\quad\quad\quad\quad\quad\quad\quad\quad\quad\quad
\times\frac{1}{2E_+(0)+2E_-(0)-E(n)-E(n^{\prime})-E(l)-E(l^{\prime})}\cr
&\quad&\quad\quad\quad\quad\quad\quad
\times\left [\frac{1}{E_+(0)+E_-(0)-E(l^{\prime})-E(n^{\prime})}+
\frac{1}{E_+(0)+E_-(0)-E(l)-E(n)}\right ]\cr
&\quad&\quad\quad\quad\quad\quad+
\frac{\langle +0|n\rangle\langle n|-0\rangle\langle -0|n^{\prime}
\rangle\langle n^{\prime}|+0\rangle\langle -0|l\rangle\langle l|+0\rangle
\langle +0|l^{\prime}\rangle\langle l^{\prime}|+0\rangle}
{[E_+(0)+E_-(0)-E(n)-E(n^{\prime})][E_+(0)+E_-(0)-E(l)-E(n)]}\cr
&\quad&\quad\quad\quad\quad\quad\quad
\times\frac{1}{E_+(0)+E_-(0)-E(l^{\prime})-E(n)}\cr
&\quad&\quad\quad\quad\quad\quad+
\frac{\langle +0|n\rangle\langle n|+0\rangle\langle -0|n^{\prime}
\rangle\langle n^{\prime}|+0\rangle\langle +0|l\rangle\langle l|-0\rangle
\langle -0|l^{\prime}\rangle\langle l^{\prime}|-0\rangle}
{[E_+(0)+E_-(0)-E(n)-E(n^{\prime})][E_+(0)+E_-(0)-E(l)-E(n^{\prime})]}\cr
&\quad&\quad\quad\quad\quad\quad\quad
\times\frac{1}{E_+(0)+E_-(0)-E(l^{\prime})-E(n^{\prime})}\biggr \}\quad ,
\label{eq: mu}
\end{eqnarray}
\begin{eqnarray}
\nu&=&-8t_{ij}t_{jk}t_{kl}t_{li}\sum_{l,l^{\prime},n,n^{\prime}=0}^{\infty}
\biggl \{ \frac{\langle +0|n\rangle\langle n|+0\rangle\langle -0|n^{\prime}
\rangle\langle n^{\prime}|-0\rangle\langle +0|l\rangle\langle l|+0\rangle
\langle -0|l^{\prime}\rangle\langle l^{\prime}|-0\rangle}
{E_+(0)+E_-(0)-E(n)-E(n^{\prime}}\cr
&\quad&\quad\quad\quad\quad\quad\quad\quad\quad\quad\quad
\times\frac{1}{2E_+(0)+2E_-(0)-E(n)-E(n^{\prime})-E(l)-E(l^{\prime})}\cr
&\quad&\quad\quad\quad\quad\quad\quad
\times\left [\frac{1}{E_+(0)+E_-(0)-E(l^{\prime})-E(n)}+
\frac{1}{E_+(0)+E_-(0)-E(l)-E(n^{\prime})}\right ]\biggr \}\quad ,
\label{eq: nu}
\end{eqnarray}
\begin{eqnarray}
\rho&=&-8t_{ij}t_{jk}t_{kl}t_{li}\sum_{l,l^{\prime},n,n^{\prime}=0}^{\infty}
\biggl \{ \frac{\langle +0|n\rangle\langle n|-0\rangle\langle -0|n^{\prime}
\rangle\langle n^{\prime}|+0\rangle\langle +0|l\rangle\langle l|-0\rangle
\langle -0|l^{\prime}\rangle\langle l^{\prime}|+0\rangle}
{E_+(0)+E_-(0)-E(n)-E(n^{\prime}}\cr
&\quad&\quad\quad\quad\quad\quad\quad\quad\quad\quad\quad
\times\frac{1}{2E_+(0)+2E_-(0)-E(n)-E(n^{\prime})-E(l)-E(l^{\prime})}\cr
&\quad&\quad\quad\quad\quad\quad\quad
\times\left [\frac{1}{E_+(0)+E_-(0)-E(l)-E(l^{\prime})}+
\frac{1}{E_+(0)+E_-(0)-E(n)-E(n^{\prime})}\right ]\cr
&\quad&\quad\quad\quad\quad\quad+
\frac{\langle +0|n\rangle\langle n|-0\rangle\langle -0|n^{\prime}
\rangle\langle n^{\prime}|+0\rangle\langle -0|l\rangle\langle l|+0\rangle
\langle +0|l^{\prime}\rangle\langle l^{\prime}|-0\rangle}
{[E_+(0)+E_-(0)-E(n)-E(n^{\prime})][E_+(0)+E_-(0)-E(l)-E(n^{\prime})]}\cr
&\quad&\quad\quad\quad\quad\quad\quad
\times\left [\frac{1}{E_+(0)+E_-(0)-E(l)-E(l^{\prime})}+
\frac{1}{E_+(0)+E_-(0)-E(l^{\prime})-E(n^{\prime})}\right ]\cr
&\quad&\quad\quad\quad\quad\quad+
\frac{\langle +0|n\rangle\langle n|-0\rangle\langle -0|n^{\prime}
\rangle\langle n^{\prime}|+0\rangle\langle +0|l\rangle\langle l|-0\rangle
\langle -0|l^{\prime}\rangle\langle l^{\prime}|+0\rangle}
{[E_+(0)+E_-(0)-E(n)-E(n^{\prime})][E_+(0)+E_-(0)-E(l)-E(n)]}\cr
&\quad&\quad\quad\quad\quad\quad\quad
\times\left [\frac{1}{E_+(0)+E_-(0)-E(l^{\prime})-E(n)}+
\frac{1}{E_+(0)+E_-(0)-E(l)-E(l^{\prime})}\right ]\biggr \}\quad .
\label{eq: rho}
\end{eqnarray}

\begin{figure}[t]
\caption{Schematic diagrams used in the determination of the effective
Hamiltonian.  The second-order diagram is plotted in (a), while the fourth-order
diagrams appear in (b-e).  The fourth-order diagrams link
two (b), three (c), or four (d)  distinct lattice sites.  The contributions
from the unlinked diagram (e) vanish. }
\end{figure}

\begin{figure}[htb]
\caption{Reduction of the bipolaron binding energy [which is the definition of
$U_c(\alpha_{anh})$] plotted as functions of the electron-phonon coupling
(a) and the anharmonicity (b).  In Fig.~2(a), three different values of the
anharmonicity parameter are chosen: $\alpha_{anh}=0.001$ (solid line);
0.01 (dotted line); and 0.1 (dashed line).  Note how the reduction of the
bipolaron binding energy increases with both $\alpha_{anh}$ and $U$.
In Fig.~2(b), the reduction in bipolaron binding energy is plotted as a
function of $\alpha_{anh}$ for three values of $|U|$: $|U|=1$ (solid line);
$|U|=4$ (dotted line); and $|U|=16$ (dashed line).  Note that even relatively
small values of $\alpha_{anh}$ produce a sharp reduction in the bipolaron
binding energy.}
\end{figure}

\begin{figure}[htb]
\caption{Charge-density-wave transition temperature at half filling for
the anharmonic Holstein model (a) or the Holstein-Hubbard model (b).
Both second-order and fourth-order approximations are plotted (the fourth-order
approximations have a peak in $T_c$).  Note that
the maximal $T_c$ for the fourth-order calculation decreases as either
the anharmonicity or Coulomb repulsion is turned on.  In Fig.~3(a),
the anharmonicity varies from $\alpha_{anh}=0.0$ (solid line), 
$\alpha_{anh}=0.001$ (dotted line), $\alpha_{anh}=0.005$ (dashed line),
and $\alpha_{anh}=0.015$ (dash-dotted line).  In Fig.~3(b), the Coulomb
repulsion is chosen to match the reduction in the bipolaron binding energy for
each value of the anharmonicity plotted in (a), i.~e., $U_c=U_c(\alpha_{anh})$
with $\alpha_{anh}=$ 0.0, 0.001, 0.005, 0.015. Note that the effect of 
anharmonicity is not easily mimicked by Coulomb repulsion except for the
smallest values of $\alpha_{anh}$.}
\end{figure}

\begin{figure}[htb]
\caption{Phase diagram of the anharmonic Holstein model (a) and the 
Holstein-Hubbard model (b) for $|U|=1.5625$ corresponding to the 
peak in the charge-density-wave transition temperature at half filling
for the harmonic Holstein model.  The solid lines denote charge-density-wave
transition temperatures and the dotted lines denote superconducting transition
temperatures.  In Fig.~4(a), the anharmonicity strength assumes four values
$\alpha_{anh}=0.0$, 0.003, 0.005, 0.01.  Note that the anharmonicity
strongly enhances superconducting solutions relative to charge-density-wave
solutions, and if $\alpha_{anh}$ is large enough, there is no CDW order.
Note furthermore, that the maximal superconducting $T_c$ is larger than the
maximal CDW $T_c$ in the harmonic case and occurs at a rather low value of
the electron filling.  In Fig.~4(b), the Coulomb repulsion
is fixed to be $U_c(\alpha_{anh})$ with $\alpha_{anh}$ the same as in (a).
Note that the effect of the Coulomb repulsion on the transition temperatures
is relatively minor, and the strong enhancement of the superconducting
$T_c$ does not occur.}
\end{figure}

\begin{figure}[htb]
\caption{Phase diagram of the anharmonic Holstein model (a) and the 
Holstein-Hubbard model (b) for $|U|=4.0$.  The solid lines denote 
charge-density-wave
transition temperatures and the dotted lines denote superconducting transition
temperatures.  In Fig.~5(a), the anharmonicity strength assumes four values
$\alpha_{anh}=0.0$, 0.01, 0.03, 0.05.  Note that the large enhancement in
the superconducting $T_c$ is seen here too (although it occurs at a larger
value of the anharmonicity). In Fig.~5(b), the Coulomb repulsion
is fixed to be $U_c(\alpha_{anh})$ with $\alpha_{anh}$ the same as in (a).
Once again, the effect of Coulomb repulsion is drastically different from
that of the anharmonicity.}
\end{figure}

\begin{figure}[htb]
\caption{Plot of the charge-density-wave--superconductor phase boundary
as functions of electron concentration and $U$.  In Fig.~6(a), the phase
diagram is plotted for the anharmonic Holstein model with $\alpha_{anh}=0.0$
(solid line), 0.001 (dotted line), 0.005 (dashed line), and 0.015 (dash-dotted
line).  In Fig.~6(b), the phase diagram is plotted for the Holstein-Hubbard
model with $U_c=U_c(\alpha_{anh})$ and $\alpha_{anh}$ chosen to have the
same values as in (a).  These figures show that the anharmonicity strongly
favors superconductivity relative to CDW order, much more so than turning
on a Coulomb repulsion.}
\end{figure}
\newpage

\begin{figure} 
\epsfxsize=4.0in
\epsffile{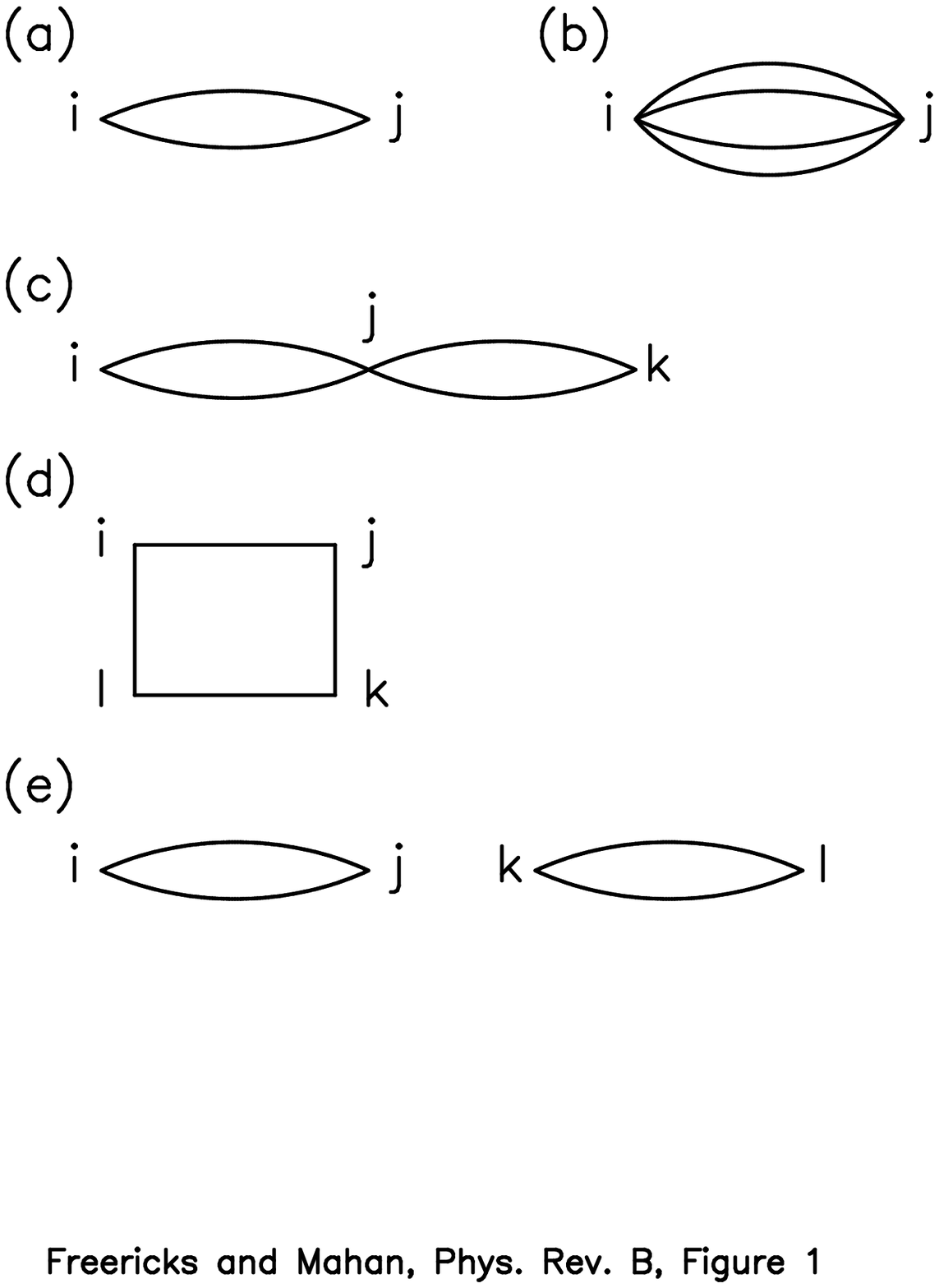}
\end{figure}

\newpage

\begin{figure} 
\epsfxsize=5.0in
\epsffile{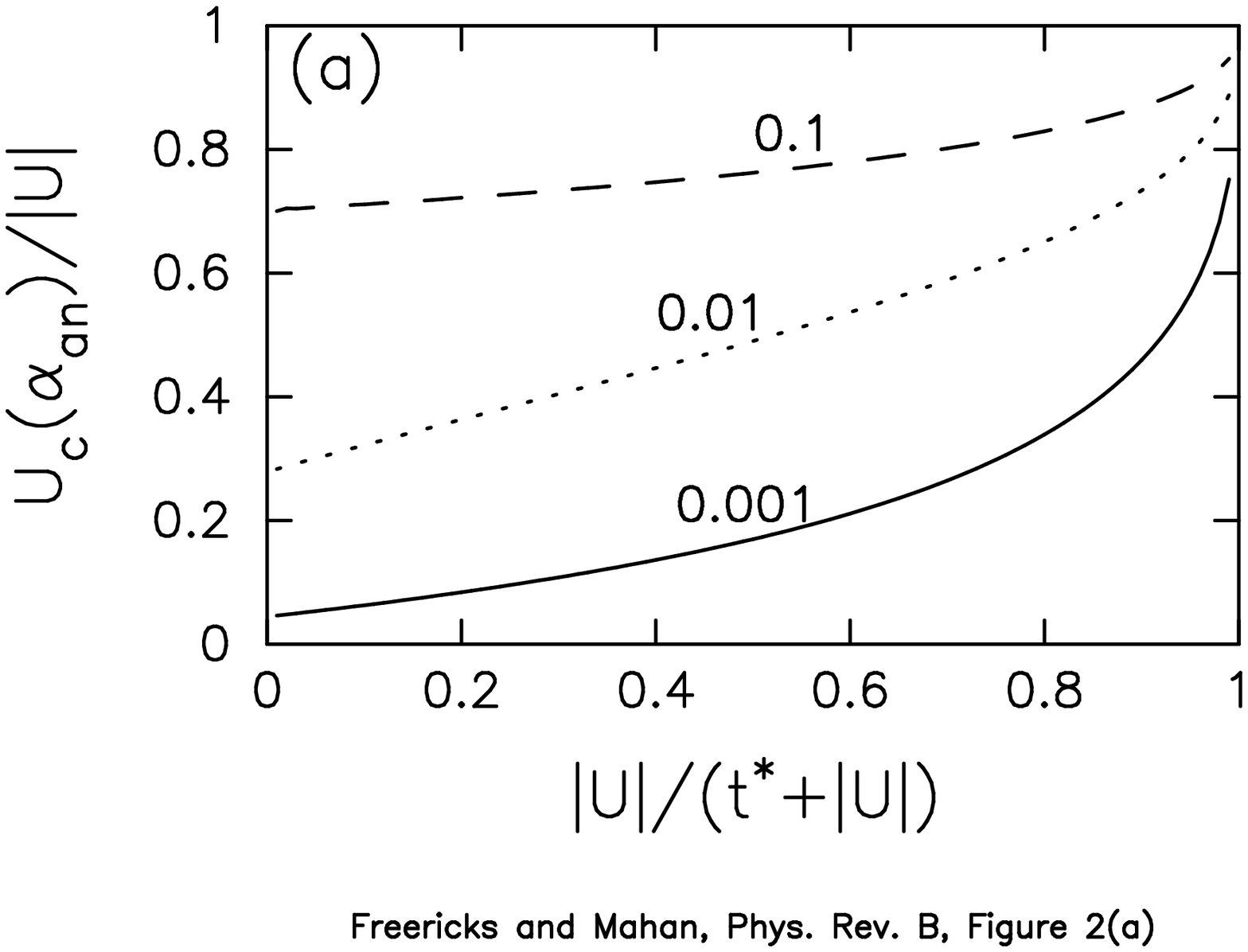}
\epsfxsize=5.0in
\epsffile{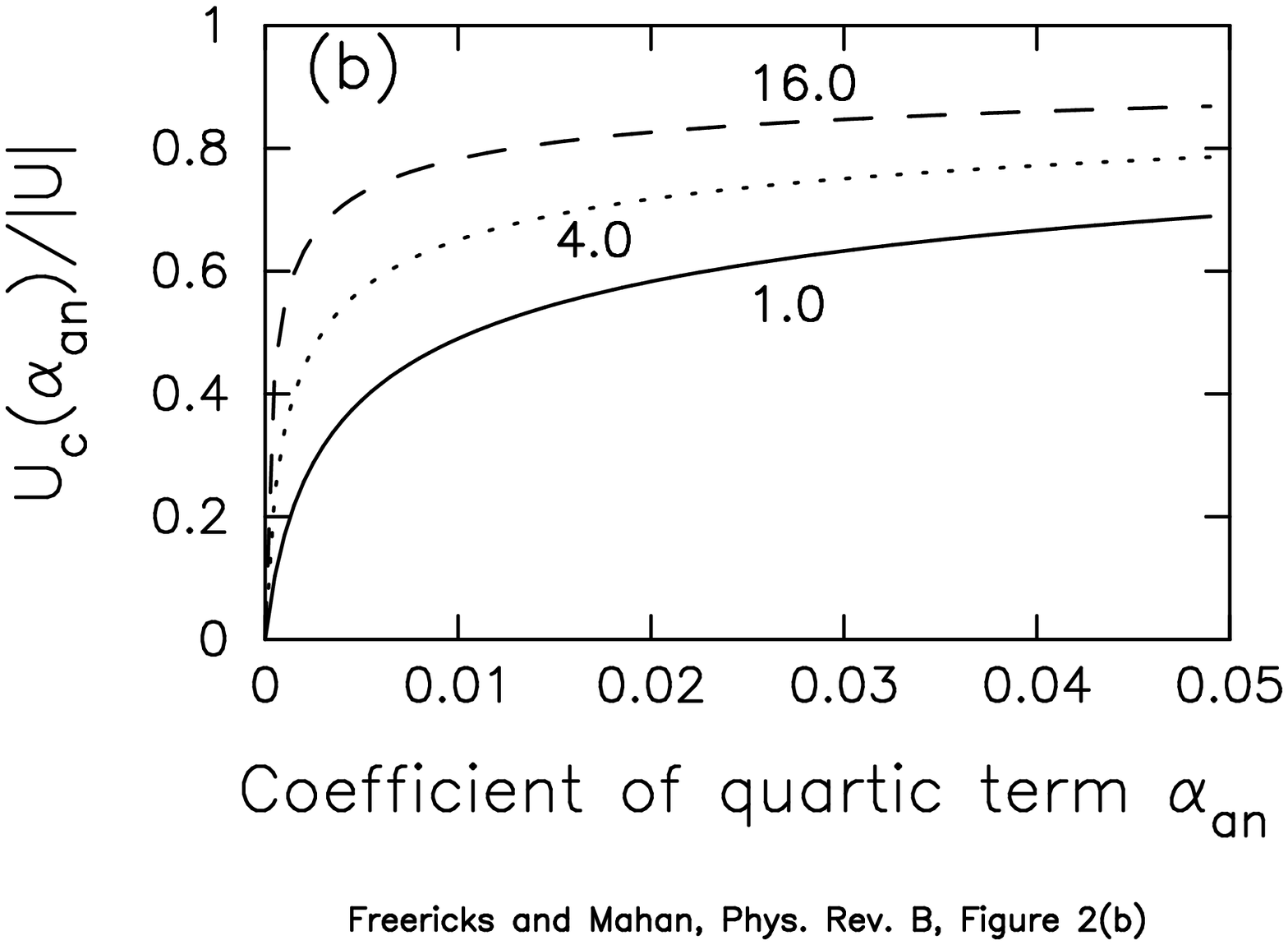}
\end{figure}

\newpage

\begin{figure} 
\epsfxsize=5.0in
\epsffile{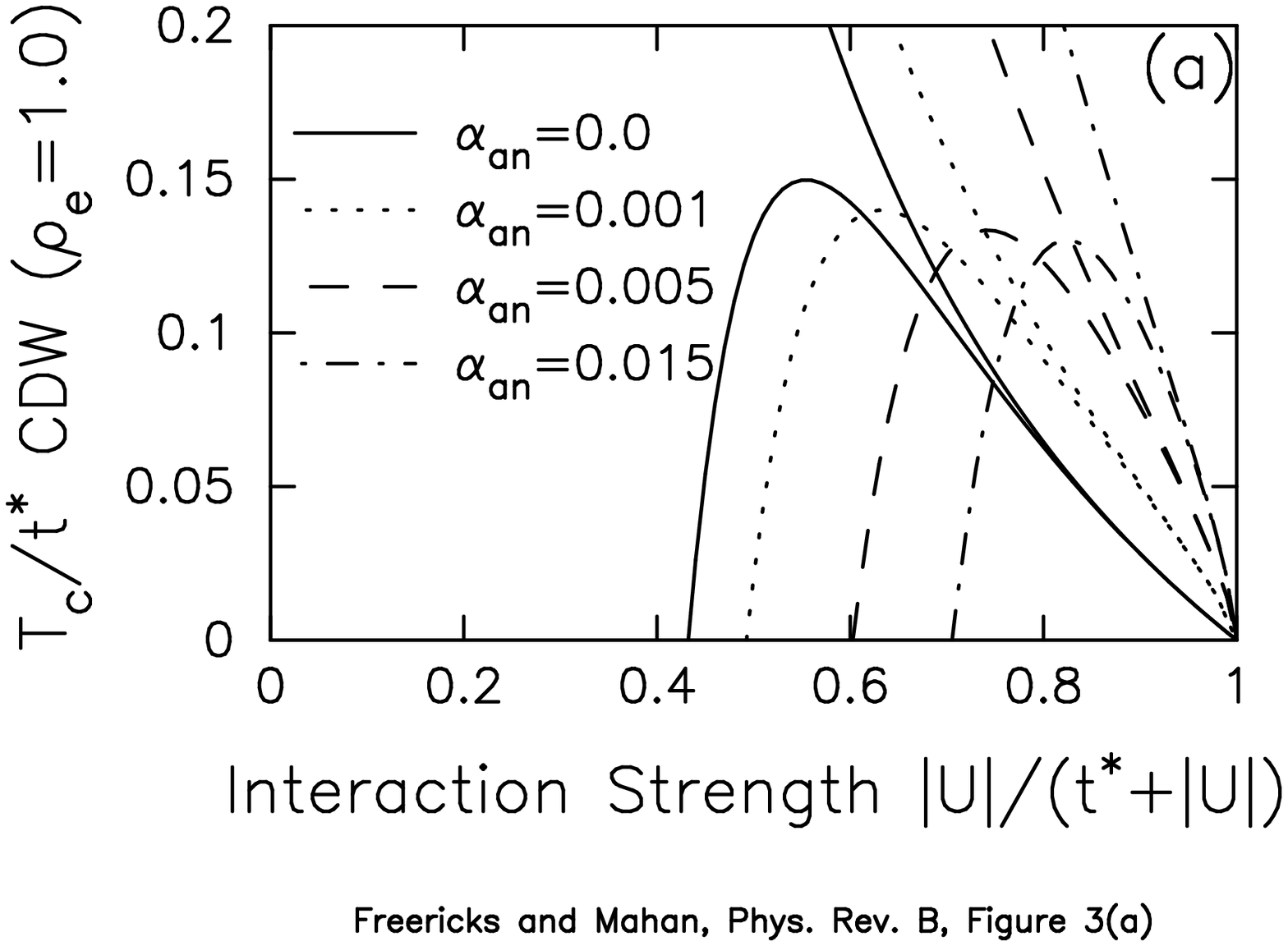}
\epsfxsize=5.0in
\epsffile{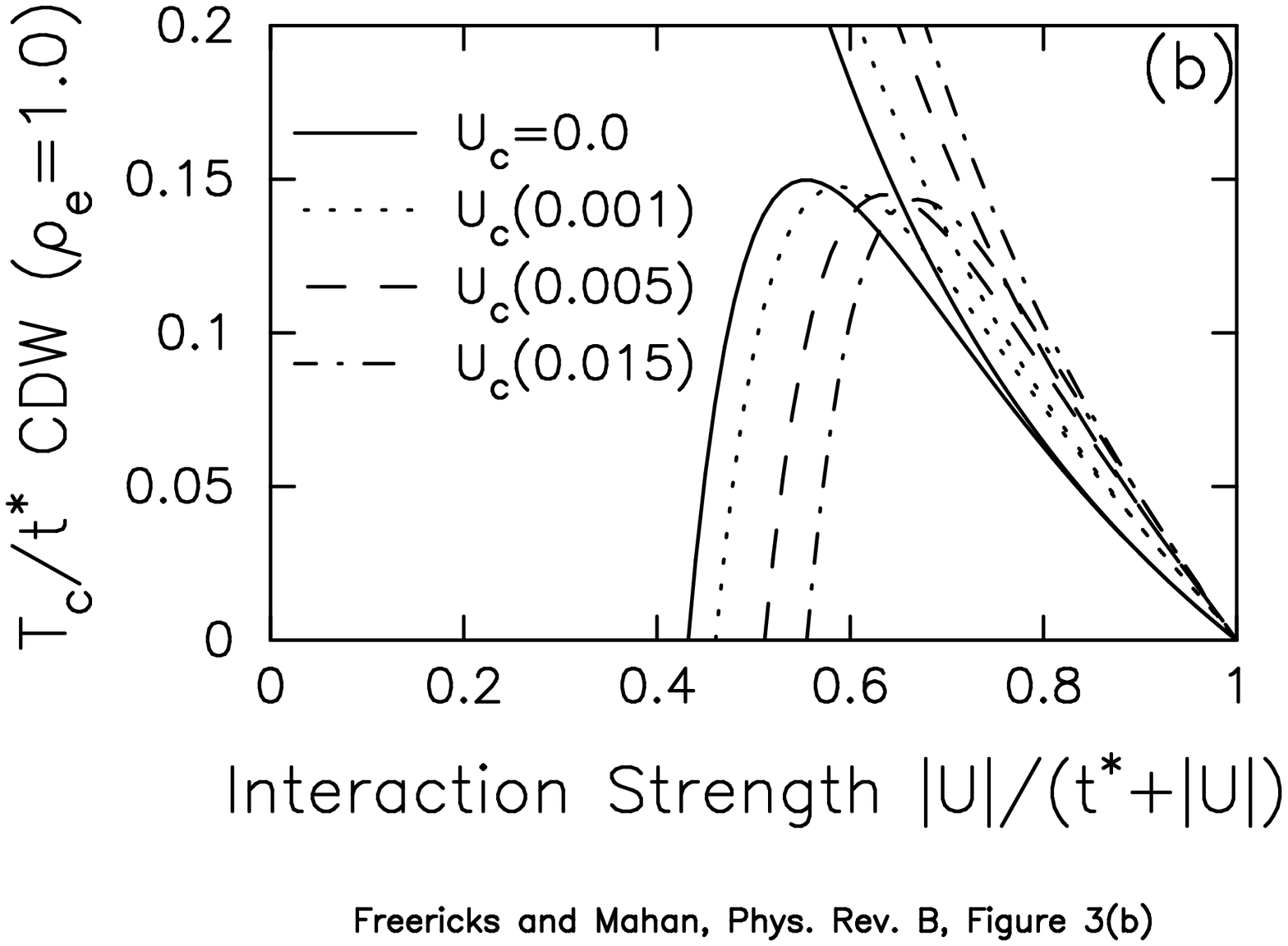}
\end{figure}

\newpage

\begin{figure} 
\epsfxsize=5.0in
\epsffile{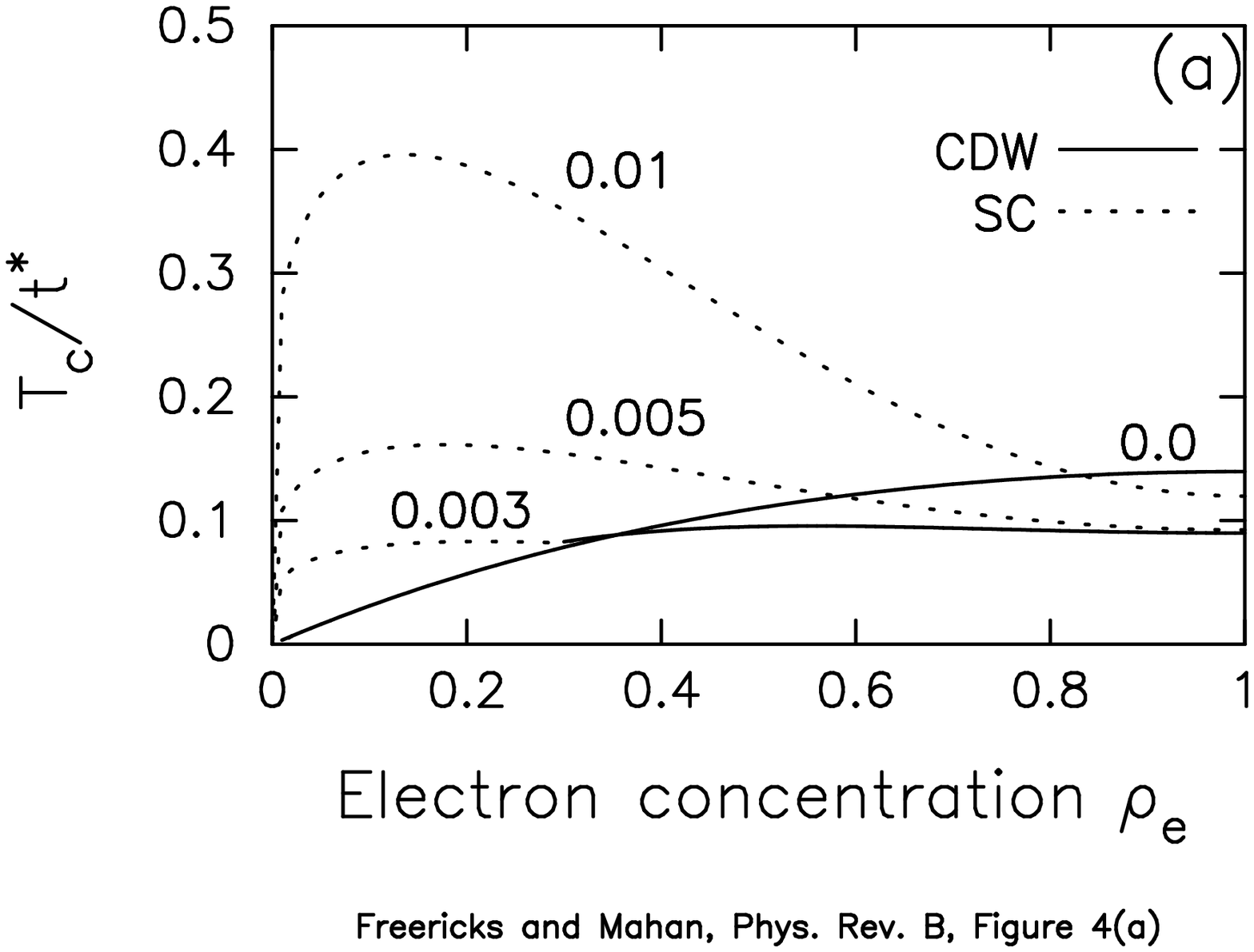}
\epsfxsize=5.0in
\epsffile{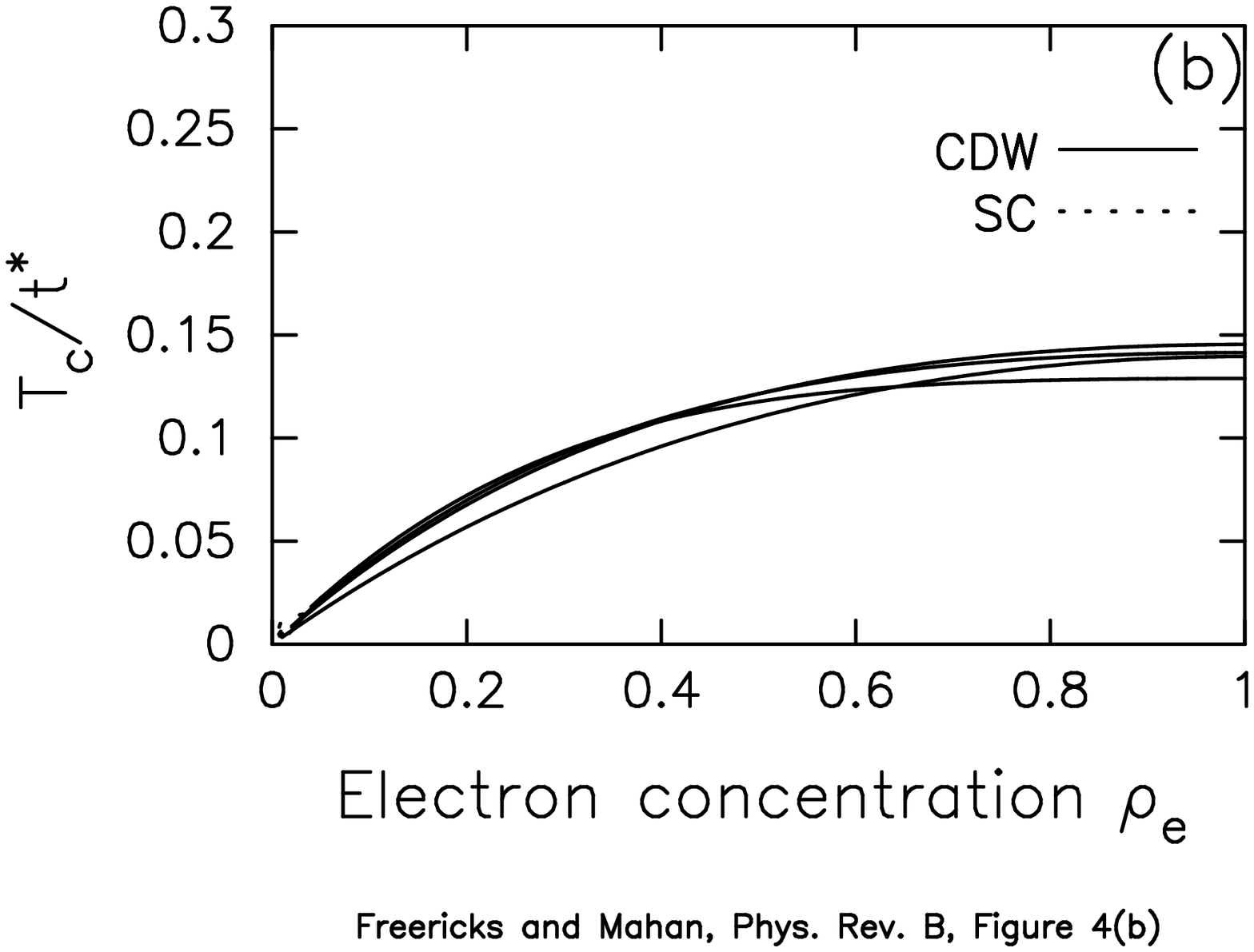}
\end{figure}

\newpage

\begin{figure} 
\epsfxsize=5.0in
\epsffile{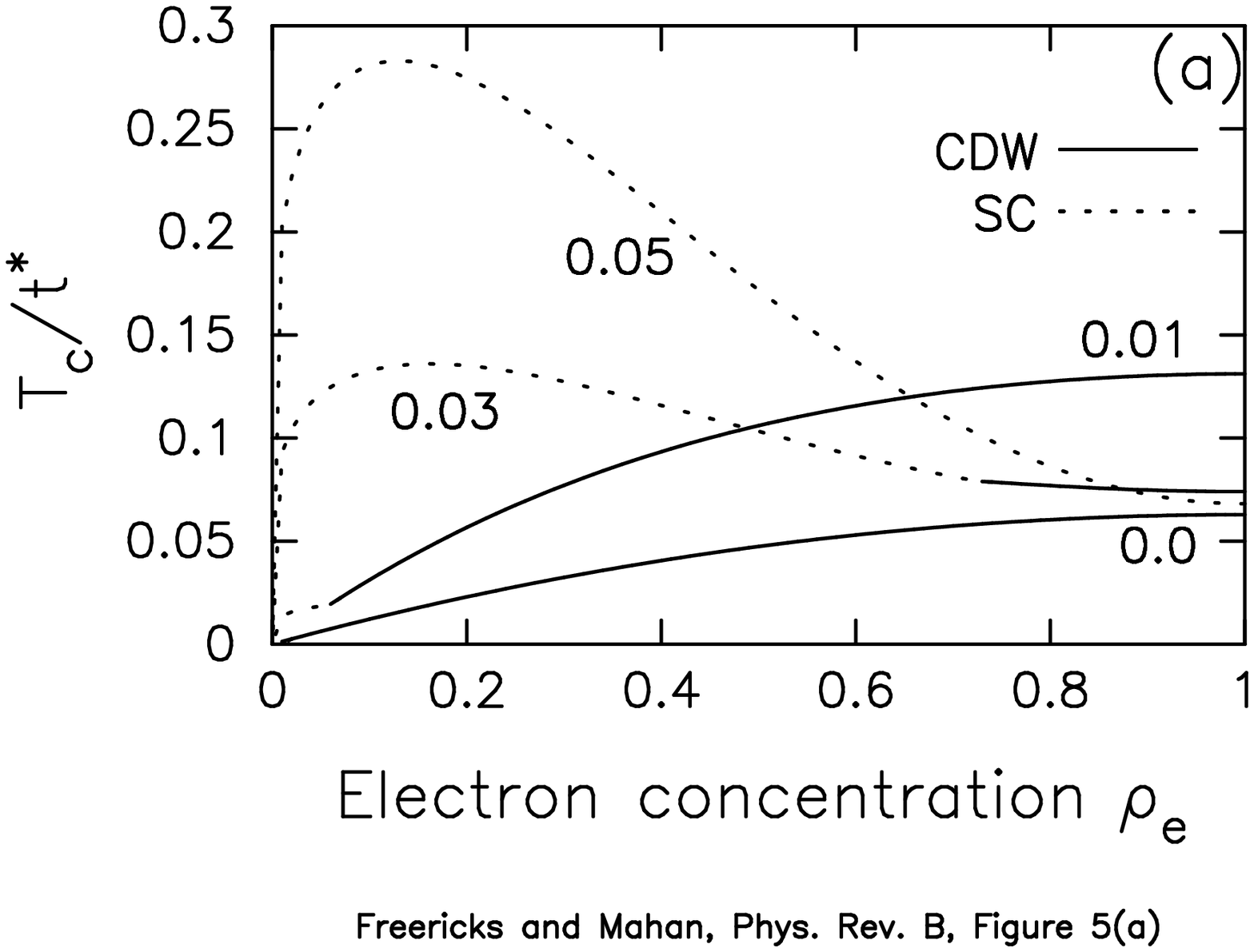}
\epsfxsize=5.0in
\epsffile{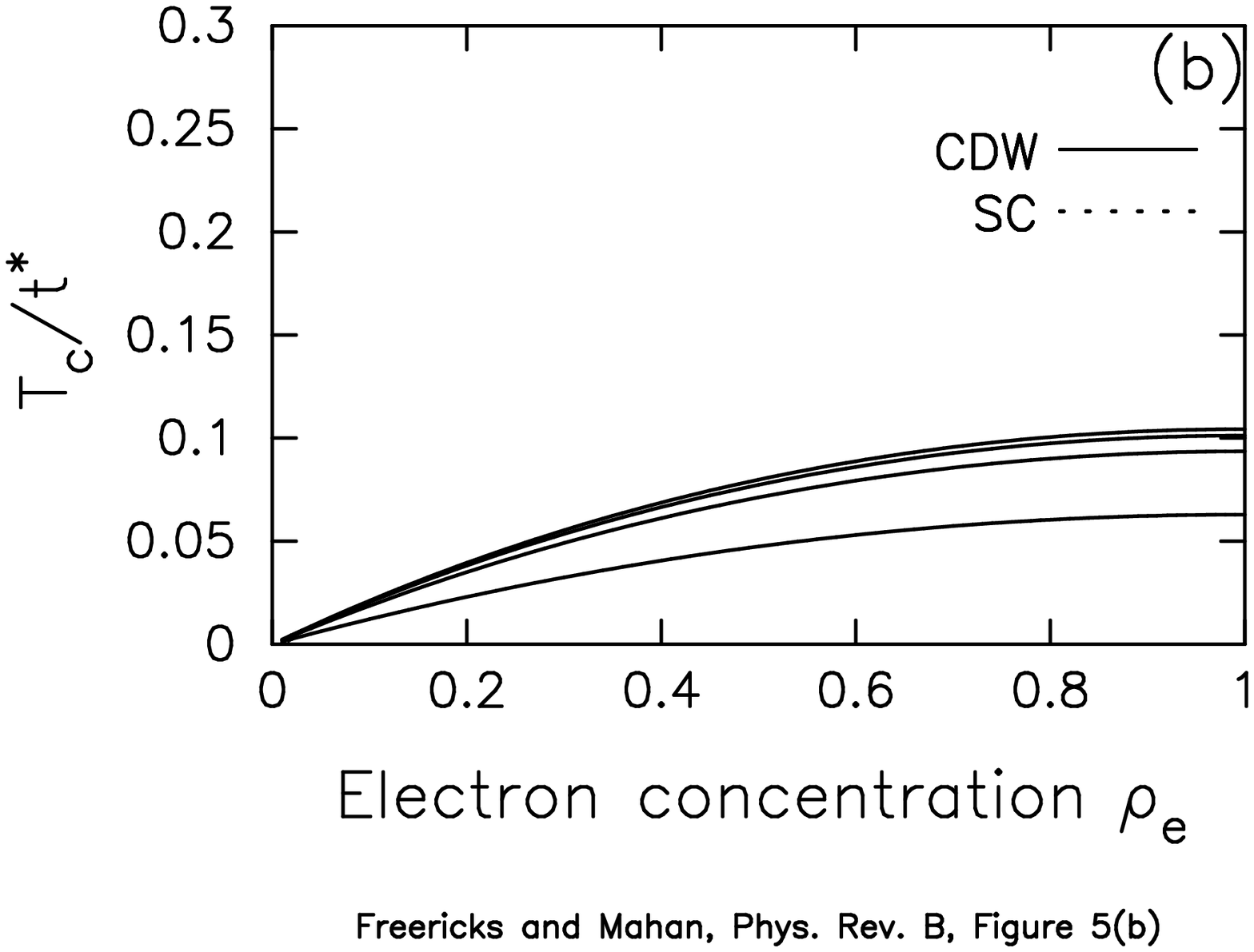}
\end{figure}

\newpage

\begin{figure} 
\epsfxsize=5.0in
\epsffile{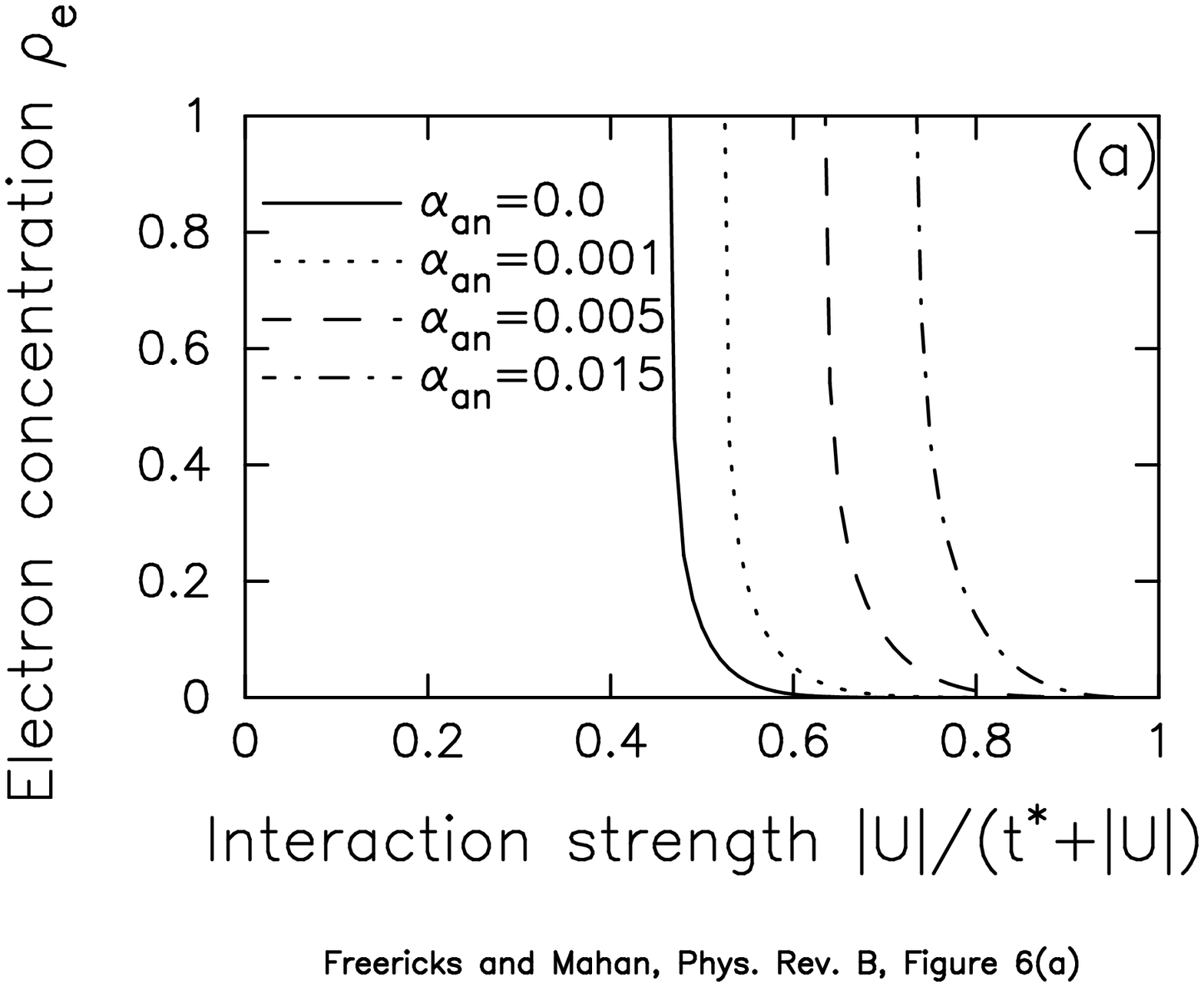}
\epsfxsize=5.0in
\epsffile{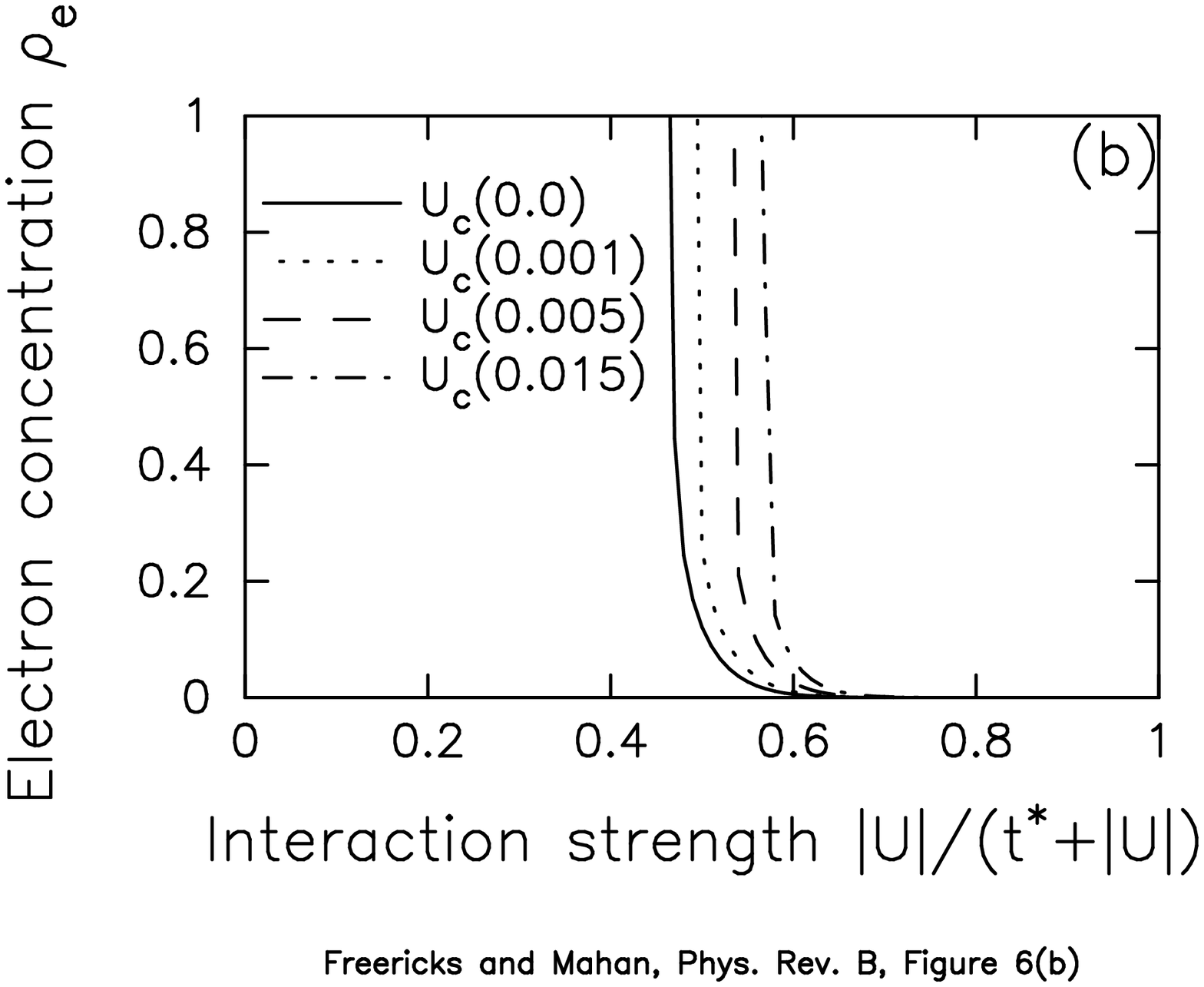}
\end{figure}

\end{document}